\documentclass[a4paper,11pt]{article}
\pdfoutput=1 %
\usepackage{jheppub} 
\usepackage[T1]{fontenc}
\usepackage{ulem}
\usepackage{feynmp-auto} 
\usepackage{adjustbox} 
\usepackage{scalerel} 
\usepackage{slashed}

\usepackage{environ}

\usepackage[frozencache,cachedir=_minted-main]{minted}
\setminted[]{fontfamily=lmtt}

\newlength{\myl}
\expandafter\let\expandafter\origequation\csname equation*\endcsname
\expandafter\let\expandafter\endorigequation\csname endequation*\endcsname
\long\def\[#1\]{\begin{equation*}#1\end{equation*}}
\RenewEnviron{equation*}{
  \settowidth{\myl}{$\displaystyle\BODY$} 
  \origequation
    \ifdim\myl>\linewidth
      \resizebox{\linewidth}{!}{$\displaystyle\BODY$}
    \else
      \BODY 
    \fi
  \endorigequation
}

\DeclareMathSymbol{\comma}{\mathpunct}{letters}{"3B} 

\usepackage{physics}
\usepackage[caption=false]{subfig}
\usepackage{graphicx}
\usepackage{caption}
\usepackage{bm}

\usepackage{array}
\usepackage{hhline}
\usepackage{booktabs}   
\usepackage{makecell} 
\renewcommand\arraystretch{1}
\usepackage{cancel}

\usepackage{multirow}

\usepackage[dvipsnames]{xcolor}

\definecolor{ForestGreen}{rgb}{0.13, 0.55, 0.13}
\definecolor{Mulberry}{rgb}{0.77, 0.29, 0.55}
\definecolor{bostonuniversityred}{rgb}{0.8, 0.0, 0.0}
\definecolor{amber}{rgb}{1.0, 0.49, 0.0}
\newcommand{\amber}[0]{\color{amber}}

\newcommand{\redBU}[0]{\color{bostonuniversityred}}

\renewcommand\arraystretch{1.0}

\title{\boldmath Exploring mixed lepton-quark interactions in non-resonant leptoquark production at the LHC}

\author[a,b,1]{João Gonçalves, \note{Corresponding author.}}
\author[a,c]{António P.~Morais,}
\author[d]{António Onofre}
\author[b]{and Roman Pasechnik}

\affiliation[a]{Departamento de F\'isica, Universidade de Aveiro and CIDMA, Campus de Santiago,\\ 3810-183 Aveiro, Portugal}
\affiliation[b]{Department of Physics, Lund
	University, \\ SE-223 62 Lund, Sweden}
\affiliation[c]{Theoretical Physics Department, CERN,\\ 1211 Geneva 23, Switzerland}
\affiliation[d]{Centro de Física da Universidade do Minho e da Universidade do Porto (CF-UM-UP),\\ 4710-057 Braga, Portugal}
\emailAdd{jpedropino@ua.pt}
\emailAdd{aapmorais@ua.pt}
\emailAdd{antonio.onofre@cern.ch}
\emailAdd{roman.pasechnik@hep.lu.se}

\abstract{Searches for new physics (NP) at particle colliders typically involve multivariate analysis of kinematic distributions of final state particles produced in a decay of a hypothetical NP resonance. Since the pair-production cross-sections mediated by such resonances are strongly suppressed by the NP scale, this analysis becomes less relevant for NP searches for masses of the BSM resonance above $1~\mathrm{TeV}$. On the other hand, $t$-channel processes are less sensitive to the mass of the virtual mediator and therefore larger phase-space can be potentially probed as well as the couplings between the NP particles and the Standard Model fields.  The fact that transitions between different generations of quarks and leptons may exist, the potential of the search presented in this article can be used, as a reference guide, to enlarge significantly the scope of searches performed at the LHC to flavour off-diagonal channels, in a theoretically consistent approach. In this work, we study non-resonant production of scalar leptoquarks which have been proposed in the literature to provide a potential avenue for radiative generation of neutrino masses, accommodating as well the existing flavour physics data. Final states involving just two muons at the LHC ($\mu^+\mu^-$), are used as a well-motivated case study.}

\begin{document} 
	\begin{flushright}
		CERN-TH-2023-072\\
		\vskip1cm
	\end{flushright}
	
	\maketitle
	\flushbottom
	
	\section{Introduction}\label{sec:intro}
	
	The continuously ongoing searches for particles beyond the Standard Model (SM) framework in resonant production channels at colliders have so far come up with negative results. Nevertheless, there are various open questions that the SM is not able to fully accommodate, such as, e.g.~dark matter, matter/anti-matter asymmetry as well as sufficient sources of CP-violation. Additionally, the SM framework can not account for neutrino masses and mixings at a renormalisable level, requiring the introduction of the 5-dimensional Weinberg operator \cite{Weinberg:1980wa}. There are however several extensions of the SM that can explain neutrino masses and mixing \cite{Zee:1985id,Gell-Mann:1979vob,Mohapatra:1979ia,Schechter:1980gr,Mohapatra:1987hh,Gu:2007ug,Foot:1988aq,Cai:2017jrq,Hagedorn:2016dze,King:2017guk,Kobayashi:2018vbk,Novichkov:2018ovf,King:2013eh}.
	
	In particular, as shown in a previous work by some of the authors \cite{Freitas:2022gqs}, the neutrino sector properties can be accommodated in an economical scalar leptoquark (LQ) model. Besides the SM particle content, it contains two additional scalar fields $S\sim(\overline{\textbf{3}}, \textbf{1}, 1/3)$ and $R\sim(\textbf{3}, \textbf{2}, 1/6)$. After electroweak symmetry breaking (EWSB), the latter gives rise to three New Physics (NP) states, one with an electric charge of $2/3e$ and two with electric charge of $1/3e$, with $|e|$ being the electron charge. LQs have been the subject of experimental searches for many years \cite{H1:1993vsn,ZEUS:1993vas,H1:1997utt,OPAL:2001fno}. However, a vast majority of these searches relies on a LQ-generation assumption, that is, a given LQ species only couples to fermions of the same generation (e.g.~a first generation LQ couples only to the electron and the up/down quarks, while the second couples only to the muon and the charm/strange quarks), which is no longer a generic assumption to take, within the models discussed nowadays. 
	\begin{figure}[htb!]
		\centering
		\includegraphics[width=0.85\textwidth,height=\textheight,keepaspectratio]{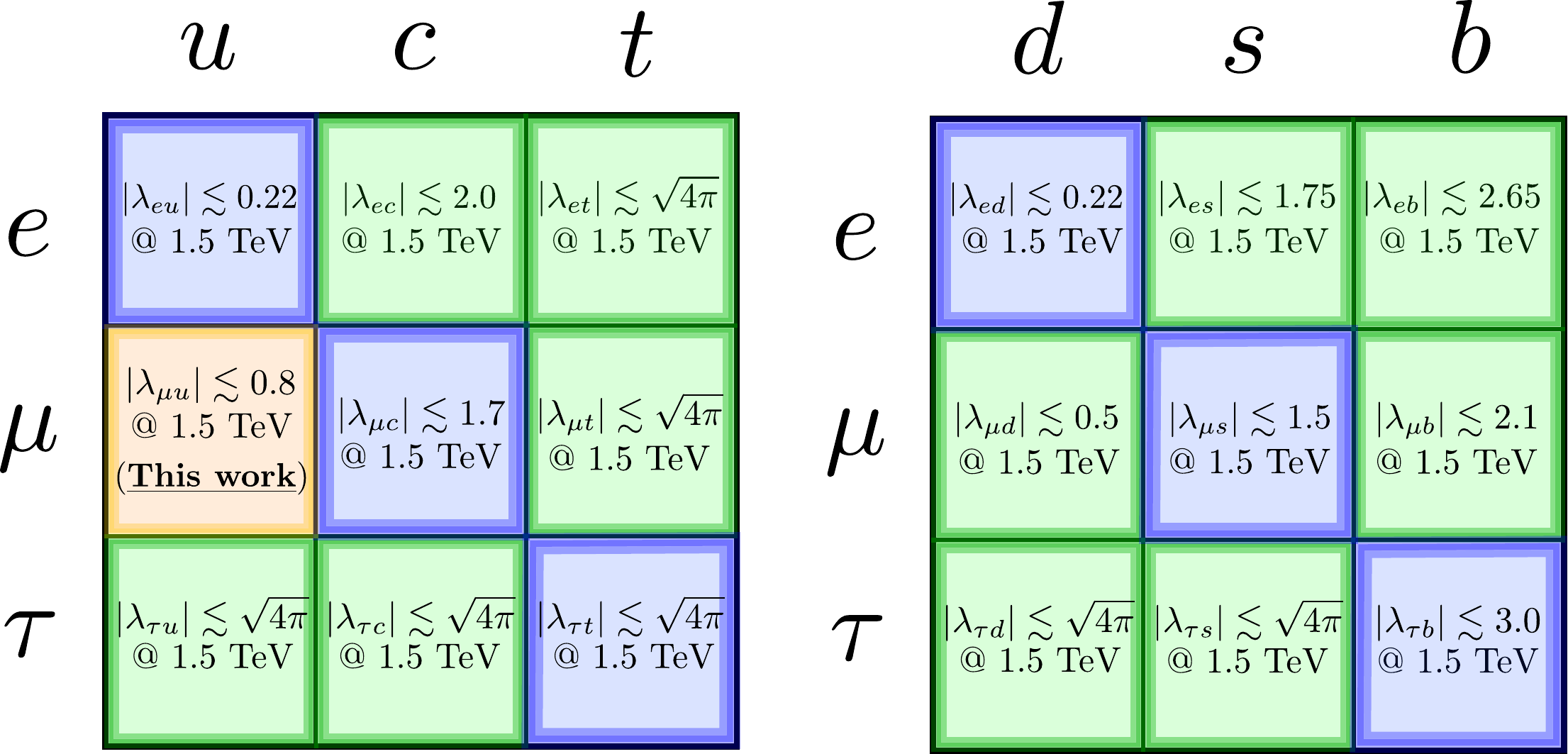}
		\caption{Allowed constraints for the LQ Yukawa couplings based on the current LHC data for a benchmark LQ mass case of $1.5~\mathrm{TeV}$. These limits take into account the experimental data from single and pair production searches, recasts from supersymmetric analyses as well as constraints from atomic parity violation, the latter of which is only applicable for couplings between electron and up/down quarks. The limits have been adapted from Ref.~\cite{Schmaltz:2018nls}.}
		\label{fig:off_diag}
	\end{figure}
	Indeed, there is still plenty of the parameter space left unexplored for off-diagonal channels (see Fig.~\ref{fig:off_diag}), with various couplings being poorly constrained, in particular, for the ones involving the first- and third-generation fermions. A detailed exploration of all possible couplings' combinations is currently lacking but provides an additional important razor on the parameter space on LQ models \cite{Babu:2020hun,Babu:2019mfe,Buonocore:2020erb} such as the one studied here.
	
	Besides, the existing LQ searches conducted at the LHC focus mostly on resonant pair-production channels, with subsequent decay of the LQs. However, pair production of heavy particles falls steeply with increasing mass for a fixed centre of mass energy. This implies that such a scaling with LQ mass directly implies that the analysis quickly looses its discriminating power as the mass approaches a TeV scale. In particular, based on the current status at the LHC, we have lower bounds of 1.5 TeV for scalar LQs and 2.0 TeV for vector LQs \cite{ATLAS:2022wcu,ATLAS:2021jyv,ATLAS:2019qpq,ATLAS:2020xov,CMS:2018ncu,CMS:2021far,CMS:2018yiq,CMS:2018txo,CMS:2018iye,CMS:2020wzx,ATLAS:2021yij}. 
	
	On the other hand, non-resonant production, with a virtual exchange of LQ in the $t$-channel, is less sensitive to the mass, with the cross-section scaling as $\lambda^2$, where $\lambda$ is the LQ Yukawa coupling to the fermions involved in the process\footnote{Do note that, even in pair-production processes, the coupling dependence still exists and becomes relevant when it is of order $\mathcal{O}(1)$, due to the presence of an additional $t$-channel contribution (see, for example, diagram PP-5 of Figure~1 of \cite{Schmaltz:2018nls}).}. In general, the existing anomalies such as the anomalous magnetic moment of the muon \cite{Muong-2:2021ojo}, the $W$-mass shift \cite{CDF:2022hxs} and the $R_{D,D^*}$ observables \cite{BaBar:2013mob, BaBar:2012obs, Belle:2015qfa, Belle:2016ure, Belle:2017ilt, LHCb:2015gmp} can be, in principle, resolved via extra contributions containing virtual NP states, with their coupling strengths to the SM fermions playing a major role. Similarly, the analysis of reactions involving virtual LQ states propagating in the $t$-channel offer an additional attractive route to further probe the model parameter space for a much larger phase space region \cite{Raj:2016aky,Bansal:2018eha,Desai:2023jxh}, while simultaneously constraining the couplings of the LQs to the SM states. 
	
	For that reason, in this work we explore collider implications of simplest and cleanest $2\to 2$ process involving the scalar LQs in the $t$-channel and the $\mu^+\mu^-$ final state motivated from the experimental point of view. It is worth mentioning that the considered model of Ref.~\cite{Freitas:2022gqs} can easily accommodate scenarios that are consistent with an extensive set of flavour observables, together with the neutrino sector constraints as well as with the exclusion bounds available from the LHC. Such benchmark scenarios give rise to $e^+e^-$, $\mu^+\mu^-$ and $\tau^+\tau^-$ final states, and in this paper we focus on the muon channel as a case study. It is also of relevance to note that flavour off-diagonal channels are also present, such as e.g. $\mu^\pm e^\mp$, $\tau^\pm \mu^\mp$ or $\tau^\pm e^\mp$, which can all be searched for. Although being outside the scope of this paper, the collider search of these final states would enable us to probe the full LQ-quark-lepton Yukawa textures.
	
	This article is organized in the following way. In Sec.~\ref{sec:model}, we introduce the LQ model, with a particular focus on its Yukawa and scalar sectors in connection to the neutrino mass and mixing properties. In Sec.~\ref{sec:collider_section}, we provide an overview of the current status of LQ searches by the CMS and ATLAS experiments and discuss the considered topology of the LQ production process, as well as the simulation methodology adopted in this work. In Sec.~\ref{sec:results} we present and discuss the numerical results of our analysis. Final remarks and conclusions are given in Sec.~\ref{sec:conclusions}.
	
	\section{The minimal scalar leptoquark model}\label{sec:model}
	
	The minimal LQ model that we consider here is inspired by a low-energy effective field theory limit of the $\mathrm{E_6}$ Grand Unification Theory extended by the family symmetry such as $\mathrm{SU(3)_F}$ or $\mathrm{SU(2)_F \times U(1)_F}$~\cite{Morais:2020ypd,Morais:2020odg}. The phenomenological viability and significance of the minimal LQ model in the context of neutrino and flavour physics have been thoroughly explored in our previous work \cite{Freitas:2022gqs}. Here, we would like to give a brief summary of the model structure and its main properties as the baseline for further phenomenological analysis of its collider implications.
	
	The considered model is based on the $\mathrm{SU(3)}\times \mathrm{SU(2)} \times \mathrm{U(1)} \times \mathbb{P}_\mathrm{B}$ symmetry group, where the so-called ``B-parity'' $\mathbb{P}_\mathrm{B}$ acts as a discrete $\mathbb{Z}_2$ symmetry. Indeed, as it was shown in Refs.~\cite{Morais:2020ypd,Morais:2020odg}, following the breaking of $\mathrm{E_6}$ to the trinification group, an accidental abelian symmetry group, $\mathrm{U(1)_W}\times \mathrm{U(1)_B}$ is generated together with the B-parity defined as
	\begin{equation}\label{eq:PB}
	\mathbb{P}_\mathrm{B} = (-1)^{2W + 2S} = (-1)^{3B + 2S} \,,
	\end{equation}
	where $S$ is the spin of the particle, and $W$ and $B$ are the charges of $\mathrm{U(1)_W}$ and $\mathrm{U(1)_B}$ symmetries. The corresponding $\mathbb{P}_\mathrm{B}$ charges are given in Tables~\ref{tab:charges_BSM} and \ref{tab:charges_SM} for the BSM and SM fields, respectively. The presence of $\mathbb{P}_\mathrm{B}$ symmetry in the low-energy limit of the theory is relevant since it forbids di-quark couplings with the scalar LQs, which are problematic since they can generate proton decay.
	
	With this being said, the model is an extension of the SM with two new scalar LQs, $S$ and $R$, with hypercharges of $1/3$ and $1/6$, respectively. The quantum numbers of the BSM particle content can be seen in Table~\ref{tab:charges_BSM}, whereas the SM states are listed in Table~\ref{tab:charges_SM}.
	\begin{table}[htb!]
		\caption{\label{tab:charges_BSM} Quantum numbers of the BSM fields. $S$ and $R$ are the scalar LQs, with the first (second) being a singlet (doublet) of $\mathbf{SU(2)_\text{L}}$.}
		\centering
		\begin{tabular}{|c|c|c|c|c|c|}
			\hline
			\textbf{Field} & $\mathbf{SU(3)_\text{C}}$ & $\mathbf{SU(2)_\text{L}}$ & $\mathbf{U(1)_\text{Y}}$ & $\mathbb{P}_\mathrm{B}$ &\textbf{\# of generations} \\ \hline
			$S$          & $\bar{\textbf{3}}$                & \textbf{1}                & $1/3$   & $-1$ & 1                          \\
			$R$            & \textbf{3}                & \textbf{2} & $1/6$ & $-1$ & 1 \\
			\hline
		\end{tabular}
	\end{table}
	\begin{table}[htb!]
		\caption{\label{tab:charges_SM} Quantum numbers of the SM fields.}
		\centering
		\begin{tabular}{|c|c|c|c|c|c|}
			\hline
			\textbf{Field} & $\mathbf{SU(3)_\text{C}}$ & $\mathbf{SU(2)_\text{L}}$ & $\mathbf{U(1)_\text{Y}}$ & $\mathbb{P}_\mathrm{B}$ &\textbf{\# of generations} \\ \hline
			$Q_\mathrm{L}$          & \textbf{3}                & \textbf{2}                & $1/6$   & $+1$& 3                          \\
			$L$            & \textbf{1}                & \textbf{2}                & $-1/2$  & $-1$& 3                          \\
			$d_\mathrm{R}$          & \textbf{3}                & \textbf{1}                & $-1/3$  & $+1$& 3                          \\
			$u_\mathrm{R}$          & \textbf{3}                & \textbf{1}                & $2/3$   & $+1$& 3                          \\
			$e_\mathrm{R}$          & \textbf{1}                & \textbf{1}                & $-1$              &$-1$ &  3 \\ 
			$H$          & $\bm{1}$                & $\bm{2}$                & $1/2$   & $+1$ & 1 \\
			\hline
		\end{tabular}
	\end{table}
	
	We define the $\mathrm{SU(2)_L}$ doublets as
	\begin{equation}\label{eq:SU2_dobs}
	Q_\mathrm{L}^i = \begin{bmatrix}
	u_\mathrm{L} \\[0.4em]
	d_\mathrm{L}
	\end{bmatrix}^i,
	\quad
	L^i = \begin{bmatrix}
	\nu_\mathrm{L} \\[0.4em]
	e_\mathrm{L}
	\end{bmatrix}^i,
	\quad
	H = \begin{bmatrix}
	H^+ \\[0.4em]
	H^0
	\end{bmatrix},
	\quad
	R = \begin{bmatrix}
	R^{2/3} \\
	R^{-1/3}
	\end{bmatrix},
	\end{equation}
	where $i=1,2,3$ is a generation index. With these symmetries in mind, we write the terms in the renormalisable Lagrangian density relevant for the current analysis. The Yukawa sector for the fermionic fields is given as
	\begin{equation}\label{eq:lagYukSM}
	\begin{aligned}
	\mathcal{L}_{\mathrm{Yuk}} = \hphantom{.}& \qty(Y_d)_{ij}  \bar{Q}^i_\mathrm{L}  d^j_\mathrm{R} H + \qty(Y_e)_{ij}  \bar{L}^i e^j_\mathrm{R} H + \qty(Y_u)_{ij}  \bar{Q}^i_\mathrm{L} u^j_\mathrm{R} \tilde{H} + {\color{red}\Theta_{ij}}  \qty(\bar{L}^{c})^i Q_\mathrm{L}^j  S \\ & + {\color{ForestGreen}\Upsilon_{ij}} \qty(\bar{e}_\mathrm{R}^c)^i u_\mathrm{R}^j  S^\dagger + {\color{blue}\Omega_{ij}} \bar{L}^i d^j_\mathrm{R} R^\dagger +  \mathrm{h.c.}\,,
	\end{aligned}
	\end{equation}
	with $\mathrm{h.c.}$ indicating Hermitian conjugate of previous terms and $\tilde{H} = i\sigma_2 H^\dagger$. All Yukawa parameters, $Y_d$, $Y_e$, $Y_u$, ${\color{red} \Theta}$, ${\color{ForestGreen} \Upsilon}$ and ${\color{blue} \Omega}$, are complex $3\times 3$ matrices\footnote{Recall that charge conjugation of a Dirac spinor is defined as $\psi^c \equiv -i\gamma^2\psi^*$, with $\gamma^2$ being the second gamma matrix. Here, $\mathrm{SU(2)}$ contractions are also left implicit. For example, $\bar{Q^c_\mathrm{L}}L \equiv \epsilon_{\alpha\beta}\bar{Q}^{c, \alpha}_\mathrm{L} L^\beta$, with $\epsilon_{\alpha\beta}$ being the Levi-Civita symbol in two dimensions.}. We follow the same notation as in \cite{Freitas:2022gqs}, where in the LQ couplings the first index is a lepton one and the second index is a quark one.
	
	The relevant terms in the scalar potential of the model $V$ can be divided into a phase-independent part, $V_1$, written as
	\begin{equation}\label{eq:scalar_potential_1}
	\begin{aligned}
	V_1 \supset \hphantom{.}& -\mu^2 \abs{H}^2 + \mu_S^2 \abs{S}^2 + \mu_R^2 \abs{R}^2  + \lambda (H^\dagger H)^2 + g_{HR} (H^\dagger H) (R^\dagger R) + \\
	&g^\prime_{HR} (H^\dagger R) (R^\dagger H) + g_{HS} (H^\dagger H)(S^\dagger S) \,,
	\end{aligned}
	\end{equation}
	and the phase-dependent part, $V_2$, given by
	\begin{equation}\label{eq:scalar_potential_2}
	\begin{aligned}
	V_2 = {\color{amber} a_{1}} R S H^\dagger + \mathrm{h.c.} \,,
	\end{aligned}
	\end{equation}
	such that $V = V_1 + V_2$.
	
	Once the Higgs doublet gains a vacuum expectation value (VEV), which in the unitary gauge corresponds to $\expval{H} = \begin{bmatrix}0 & (v + h)/\sqrt{2} \end{bmatrix}^T$ and $v \approx 246~\mathrm{GeV}$, the mass for the Higgs field remains identical to that in the SM, $m_h^2 = 2\lambda v^2$. The $S$ LQ and the second component of the $R$ LQ doublet mix (corresponding to the LQs with an electric charge of $1/3e$), which leads to the squared mass matrix,
	\begin{equation}\label{eq:LQ_13}
	M^2_{LQ^{1/3}} = \begin{bmatrix}
	\mu_S^2 + \dfrac{g_{HS}v^2}{2} & \dfrac{v {\color{amber} a_1}}{\sqrt{2}} \\
	\dfrac{v {\color{amber} a_1}}{\sqrt{2}} & \mu_R^2 + \dfrac{G v^2}{2}
	\end{bmatrix} \,,
	\end{equation}
	where we have assumed that ${\color{amber} a_1}$ is real, and $G = (g_{HR} + g^\prime_{HR})$. The eigenvalues read
	\begin{equation}\label{eq:eigenvalues_LQ_13}
	\begin{aligned}
	m^2_{S_1^{1/3}} = \frac{1}{4} \Bigg(2\mu_R^2 + 2\mu_S^2 + v^2(G + g_{HS}) - \sqrt{(2\mu_R^2 - 2\mu_S^2 + (G - g_{HS})v^2)^2 + 8\abs{{\color{amber} a_1}}^2v^2}\Bigg) \,,\\
	m^2_{S_2^{1/3}} = \frac{1}{4} \Bigg(2\mu_R^2 + 2\mu_S^2 + v^2(G + g_{HS})  + \sqrt{(2\mu_R^2 - 2\mu_S^2 + (G - g_{HS})v^2)^2 + 8\abs{{\color{amber} a_1}}^2v^2}\Bigg) \,, \\
	\end{aligned}
	\end{equation}
	where we adopt the notation for the mass eigenstates of $S^{1/3}_1$ and $S^{1/3}_2$. Do note that one can diagonalise the matrix in Eq.~\eqref{eq:LQ_13} via an unitary transformation, that is,
	\begin{equation}\label{eq:biunitary_LQ^(1/3)}
	M^{\mathrm{diag}}_{LQ^{1/3}} = Z^H M^2_{LQ^{1/3}} Z^{H, \dagger} \,,
	\end{equation}
	where $Z^H$ is a unitary matrix and $M^{\mathrm{diag}}_{\mathrm{LQ}^{1/3}}$ is the $2\times 2$ LQ mass matrix in the diagonal form. So the mixing is parameterized by a single angle $\theta$, which in terms of the physical LQ masses reads
	\begin{equation}\label{eq:mix_angle}
	\sin(2\theta) = \frac{\sqrt{2}v {\color{amber} a_1}}{m_{S^{1/3}_1}^2 - m_{S^{1/3}_2}^2} \,.
	\end{equation}
	The third LQ state with $2/3 e$ charge is denoted as $S^{2/3}$ and has a tree-level mass,
	\begin{equation}\label{eq:eigenvalues_LQ_23}
	m^2_{S^{2/3}} = \mu_R^2 + \frac{g_{HR} v^2}{2} \,.
	\end{equation}
	
	A similar analysis can be conducted in both the quark and lepton sectors. Since there are no new quarks or leptons in the considered model, the fermion sector can be made identical to that in the SM. In particular, for simplicity of the forthcoming numerical analysis, we consider the basis where both the lepton and the up-quark sectors are mass-diagonal, in such a way that the Cabibbo–Kobayashi–Maskawa (CKM) matrix is fully in the down-quark sector and the Pontecorvo–Maki–Nakagawa–Sakata (PMNS) matrix is fully present in the neutrino sector. Taking this into account, the quark mass matrices can be cast as 
	\begin{equation}\label{eq:massmatrices_quarks}
	M_u = \frac{v}{\sqrt{2}} \begin{bmatrix} 
	(Y_u)_{11} & 0 & 0 \\[0.2em]
	0 & (Y_u)_{22} & 0 \\[0.2em]
	0 & 0 & (Y_u)_{33}
	\end{bmatrix},
	\quad
	M_d = \frac{v}{\sqrt{2}} \begin{bmatrix} 
	(Y_d)_{11} & (Y_d)_{12} & (Y_d)_{13} \\[0.2em]
	(Y_d)_{21} & (Y_d)_{22} & (Y_d)_{23} \\[0.2em]
	(Y_d)_{31} & (Y_d)_{32} & (Y_d)_{33}
	\end{bmatrix},
	\end{equation}
	whose diagonalisation results in the SM quarks and can performed in the standard way,
	\begin{equation}\label{eq:biunitary_quarks}
	M^{\mathrm{diag}}_{u} = M_{u}\,, \quad M^{\mathrm{diag}}_{d} = U_{d,\mathrm{L}} M_{d} U_{d,\mathrm{R}}^\dagger,
	\end{equation}
	such that the CKM matrix is given as $V = U_{d,\mathrm{L}}$. The charged lepton matrix can be written as
	\begin{equation}\label{eq:charged_lepton}
	M_e = \frac{v}{\sqrt{2}} \begin{bmatrix}
	(Y_e)_{11} & 0 & 0  \\[0.5em]
	0 & (Y_e)_{22} & 0  \\[0.5em]
	0 & 0 & (Y_e)_{33}
	\end{bmatrix}\,.
	\end{equation}
	In a similar vein to the LQ sector, we can invert these relations such that the physical parameters (masses and mixings) can be used as input. Namely, we can write 
	\begin{equation}\label{eq:inverted_eqs}
	\begin{aligned}
	Y_d = \frac{\sqrt{2}}{v} V^\dagger M_d^\mathrm{diag}, \quad\quad Y_u = \frac{\sqrt{2}}{v} M_u^\mathrm{diag}, \quad\quad Y_e = \frac{\sqrt{2}}{v} M_e^\mathrm{diag} \,.
	\end{aligned}
	\end{equation}
	Here, we have assumed that the down-quark right-handed mixing matrix is equal to the identity matrix. Since there are no right-handed neutrino states in the considered minimal LQ model, light neutrinos can acquire their small masses only radiatively. Indeed, the presence of LQs leads to one-loop diagrams that contribute to the neutrino masses. The latter are generated through the mixing between the $S$ and $R$ doublet, via the term ${\color{amber} a_1} R S H^\dagger$ in Eq.~\eqref{eq:scalar_potential_2}. The corresponding one-loop mass form can be illustrated as \cite{Dorsner:2017wwn,Zhang:2021dgl,AristizabalSierra:2007nf,Pas:2015hca,Cai:2017jrq,Cata:2019wbu},
	\begin{equation}\label{eq:loop_function_neutrinos}
	\qty(M_{\nu})_{ij} = \adjincludegraphics[valign=c, height=1.8cm, raise=2.25\baselineskip]{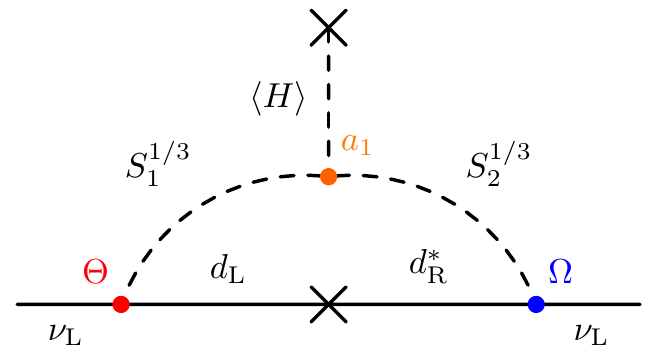}\,,
	\end{equation}
	whose loop function will depend on the ${\color{blue} \Omega}$ and ${\color{red} \Theta}$ Yukawa matrices, ${\color{amber} a_1}$ trilinear coupling and squared mass ratios between the particles running in internal lines of the loop. The one-loop integral can be analytically evaluated. In the limit of $m_{\mathrm{LQ}} \gg m_d$, this corresponds to
	\begin{equation}\label{eq:loop_integral}
	(M_\nu)_{ij} = \frac{N_c}{16\pi^2(m_{S^{1/3}_2}^2 - m_{S^{1/3}_1}^2)}\frac{v {\color{amber} a_1}}{\sqrt{2}}\ln(\frac{m_{S^{1/3}_2}^2}{m_{S^{1/3}_1}^2})\sum_{n,m} \qty(m_d)_n V_{nm} \qty({\color{red}\Theta_{im}}{\color{blue}\Omega_{jn}} + {\color{red}\Theta_{jm}}{\color{blue}\Omega_{in}}),
	\end{equation}
	where $n,k,m=1,2,3$, $N_c$ is the number of colours, i.e. $N_c=3$ in QCD, and $V$ is the CKM mixing matrix. Identical formulas to those in literature \cite{Dorsner:2017wwn,Zhang:2021dgl,AristizabalSierra:2007nf,Pas:2015hca,Cai:2017jrq,Cata:2019wbu} can be obtained by using the relation \eqref{eq:mix_angle}. Note that in the limit of no mixing (${\amber a_1} \rightarrow 0$) the neutrinos remain massless in the current minimal setup. With these new loop-generated contributions, the zero entries of the mass matrix are filled and non-zero masses and mixing for all three light active neutrinos are generated yielding a viable neutrino phenomenology and simultaneously constraining the parameters in the LQ sector. The methodology of the fit is described in our previous work \cite{Freitas:2022gqs}.
	
	\section{Non-resonant production of leptoquarks at colliders}\label{sec:collider_section}
	
	At a TeV scale, searches for LQs at both ATLAS and CMS have been extensively conducted over the past years \cite{ATLAS:2022wcu,ATLAS:2021jyv,ATLAS:2019qpq,ATLAS:2020xov,CMS:2018ncu,CMS:2021far,CMS:2018yiq,CMS:2018txo,CMS:2018iye,CMS:2020wzx,ATLAS:2021yij}. These searches are in part driven by the fact that LQs are coloured particles, with a favoured production in hadron-hadron collisions via fusion of quarks and gluons. Besides, such searches were stimulated by recent anomalous results in the flavour sector, such as the $R_{D,D^*}$ anomaly \cite{BaBar:2013mob, BaBar:2012obs, Belle:2015qfa, Belle:2016ure, Belle:2017ilt, LHCb:2015gmp}, as well as anomalous results in B-meson decays in the muon channel \cite{LHCb:2020lmf, LHCb:2020gog, LHCb:2021xxq}. Although a more recent measurement of $R_{K,K^*}$ stands in consistency with the SM predictions \cite{LHCb:2022zom}, the LQ phenomenology still remains a hot topic in the searches for NP at hadron colliders.
	
	The most recent experimental searches are summarised in Table~\ref{tab:CMS_ATLAS}. From here, we note that the searches involving third-generation fermions (tau, bottom and top) appear to be dominant across both ATLAS and CMS \cite{CMS:2020wzx, ATLAS:2021yij, ATLAS:2021oiz, ATLAS:2020dsf, ATLAS:2019qpq, CMS:2018iye, CMS:2018txo}. Indeed, the flavour anomalies in the muon sector seem to favour high values for the couplings between the second and third generations. However, as already noted earlier on in the introduction, a vast majority of these searches relies on the LQ-generation assumption (i.e.~with flavour-diagonal LQ-quark-lepton textures). Although, in this regard, we do note that the more recent searches at ATLAS are now looking for such transition patterns, with the focus on 2nd/3rd generation transitions \cite{ATLAS:2022wcu, ATLAS:2020xov, ATLAS:2020dsk, ATLAS:2021mla}. Besides this, most searches seem to tackle pair-production channels, which in principle are model independent, since the corresponding cross-section rates depend on the strong coupling constant, the LQ mass and the branching ratios (BRs). The BRs, however, can be controlled via a free parameter ($\beta$), such that the coupling between leptons and quarks (neutrinos and quarks) is given by $\sqrt{\beta}\lambda$ ($\sqrt{1-\beta}\lambda$), where $\lambda$ is a coupling constant.
	
	\begin{figure}[htb!]
		\centering
		\includegraphics[width=0.33\textwidth,height=\textheight,keepaspectratio]{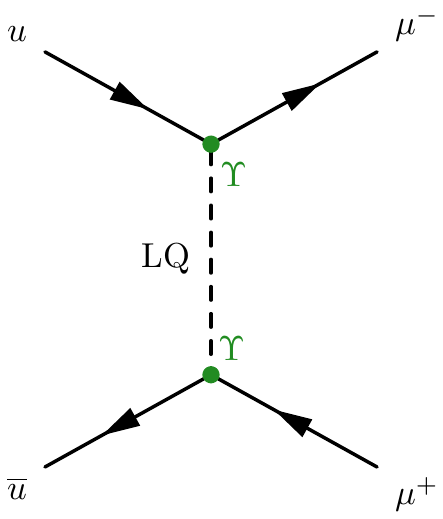}
		\caption{Topology for the non-resonant production of scalar LQs with a muon pair in the final state. The initial state $u\bar u$ is any allowed combination of up-type quarks coming from the colliding protons.}
		\label{fig:feyn_diagrams}
	\end{figure}
	
	However, pair-production channels are not strongly sensitive to the LQ coupling to fermions, whose numerical size can have a significant impact on flavour observables. As mentioned above, the cross-section for pair-production falls steeply with increasing mass. Therefore we propose to study the impact of the scalar LQ on the pair production of two leptons as this is more sensitive to the coupling structure and allows to reach a higher mass range as discussed below.
	Namely, we plan to study non-resonant production (via the $t$-channel exchange) of scalar LQ which is more sensitive to the coupling and less sensitive to the mass of the LQ, although, as we will later note, the mass of the LQ can still indirectly impact the observables through either angular or kinematic distributions. 
	\begin{table}[htb!] \centering
		\caption{A summary of the most recent searches for scalar-type LQs in the past years, by both the ATLAS and CMS experiments at the LHC. $\bm{S^{1/3}}$ is a LQ with $1/3e$ electric charge and $\bm{S^{2/3}}$ is a LQ with $2/3e$ electric charge. We denote by $\bm{S}$ the searches where the electric charge of the LQ is not relevant and/or not specified. Here, ``BR'' indicates the branching ratio.}
		\label{tab:CMS_ATLAS}
		\begin{small}
			\hspace*{-0.5em}
			\begin{tabular}{|@{}ccccc@{}|}\hline
				\textbf{Scalar} & \textbf{Decay} & \textbf{Mass (GeV)} & \textbf{Comments} & \textbf{Refs.} \\ \hline
				\multirow{7}{*}{\textbf{$\bm{S^{1/3}}$}} & $S^{1/3}\rightarrow t\ell/b\nu$ & \makecell{$[300, 1440]$} & \makecell{$\ell = e,\mu$. BR-dependent exclusions \\ also provided. Vector LQs also studied.} & \cite{ATLAS:2022wcu} \\[0.5em] \cline{2-5}
				& $S^{1/3}\rightarrow t\tau/b\nu$ & \makecell{$[400, 1250]$ \\ $[200, 800]$} & \makecell{BR-dependent exclusions also \\ provided. Vector LQs studied in \cite{ATLAS:2021jyv}.} & \makecell{\cite{ATLAS:2021jyv} \\ \cite{ATLAS:2019qpq}} \\[0.5em] \cline{2-5} 
				& $S^{1/3}\rightarrow t\ell$ & $[900, 1470]$ & \makecell{$\ell = e,\mu$. BR-dependent \\ exclusions also provided.} & \cite{ATLAS:2020xov} \\[0.5em] \cline{2-5} 
				& $S^{1/3}\rightarrow ue/\nu d$ & $[200, 1435]$ & \makecell{BR-dependent exclusions also provided. \\ Assumes 1st-generation LQ.} & \cite{CMS:2018ncu} \\[0.5em] \hline \hline
				\multirow{11}{*}{\textbf{$\bm{S^{2/3}}$}} &  $S^{2/3}\rightarrow u\nu_e$ & $[1000, 2000]$ & \makecell{Single production with coupling dependent \\ exclusions. Assumes 1st-generation LQ.} & \cite{CMS:2021far}\\[0.5em]  \cline{2-5}
				& $S^{2/3}\rightarrow \mu s$ & $[800, 1500]$ & \makecell{Exclusions depend on the mass of a \\ dark matter particle. 2nd-generation LQ.} & \cite{CMS:2018yiq}\\[0.5em]  \cline{2-5}
				& $S^{2/3}\rightarrow b \tau$ & \makecell{$[250, 1020]$ \\ $[200, 740]$} & \makecell{BR-dependent exclusions also \\ provided. Assumes 3rd-generation LQ. \\ Single production is considered in \cite{CMS:2018txo}} & \makecell{\cite{CMS:2018iye} \\ \cite{CMS:2018txo}}\\[0.5em]  \cline{2-5}
				& $S^{2/3}\rightarrow t\nu/b\ell$ & $[300, 1390]$ & \makecell{$\ell = e,\mu$. BR-dependent exclusions also \\ provided. Vector LQs also studied.} & \cite{ATLAS:2022wcu}\\[0.5em]  \cline{2-5}
				& $S^{2/3}\rightarrow t\nu/b\tau$ & \makecell{$[980, 1730]$ \\ $[400, 1260]$ \\ $[400, 1220]$ \\ $[400, 1240]$ \\ $[200, 800]$} & \makecell{BR-dependent exclusions also \\ provided. Assumes 3rd-generation LQ.} & \makecell{\cite{CMS:2020wzx} \\ \cite{ATLAS:2021yij} \\ \cite{ATLAS:2021oiz} \\ \cite{ATLAS:2020dsf} \\ \cite{ATLAS:2019qpq}} \\[0.5em] \hline \hline
				\multirow{7}{*}{\textbf{$\bm{S}$}} & $S\rightarrow q\ell$ & $[400, 1800]$ & \makecell{$\ell = e,\mu$ and $q = u, d, c, s, b$. \\ BR-dependent exclusions also provided.} & \cite{ATLAS:2020dsk}\\[0.5em]  \cline{2-5}
				& $b\bar{s}\rightarrow \ell^+\ell^-$ & $< 2400$ & \makecell{$\ell = e,\mu$. Constraints on four-fermion current. \\ Constraint in terms of mass/coupling ratio.} & \cite{ATLAS:2021mla}\\[0.5em] \cline{2-5}
				& $S \rightarrow \tau\nu$ & \makecell{$< 5900$ \\ democratic} & \makecell{LQ exchange in the $t-\mathrm{channel}$. Constraints \\ strongly depend on couplings (see Table~1) } & \cite{CMS:2022krd}\\[0.5em] \cline{2-5}
				& $S \rightarrow \mu j$ & $[200, 1530]$ & \makecell{$j$ is a light jet (up or down). BR \\ dependent constraints also provided} & \cite{CMS:2018lab}\\[0.5em] \hline \hline
			\end{tabular}
		\end{small}
	\end{table}
	
	The signal we propose to tackle is shown in Figure~\ref{fig:feyn_diagrams} which corresponds to the $t$-channel process with a muon/anti-muon pair in the final state. We have developed a specific \verb|UFO| model \cite{Degrande:2011ua} for the signal using the \verb|SARAH| package \cite{Staub:2013tta} which allows for the numerical implementation of the model's Lagrangian as well as the necessary computations of the model's interactions and mass spectra. The \verb|UFO| model was linked to the Monte-Carlo generator \verb|MadGraph| \cite{Alwall:2014hca} to compute leading-order (LO) matrix elements for the signal. Detailed information on the signal \texttt{UFO} model (\texttt{LQ\_model\_py3}) can be found in \url{https://github.com/Mrazi09/LQ_collider_project}.
	
	The main backgrounds for this signal were also generated with \texttt{MadGraph} and include the ones from $\mathrm{Z^0+jets}$ (with up to 4 additional jets), $t\bar{t}\mathrm{+jets}$ (with up to 3 jets) and di-boson plus jets (which encapsulates $WW$, $\mathrm{Z^0Z^0}$ and $\mathrm{Z^0}W$ production, and we consider up to 3 additional jets). For the simulation of proton-proton collisions at the LHC, we follow standard methodologies in the literature. The events are interfaced with \verb|Pythia8| for hadronization and showering, followed by fast detector simulation using \verb|Delphes| \cite{deFavereau:2013fsa}, where we have considered the ATLAS detector. Kinematic and angular data are extracted from the \verb|ROOT| files generated by \verb|Delphes| \cite{Brun:1997pa}.
	
	The rates of background processes beyond LO are well-known in literature, and as such, we re-weight our events based on the known values \cite{Catani:2009sm,Balossini:2009sam,Muselli:2015kba,Campbell:2011bn}. For the signal, however, we take the LO cross-section as given in \verb|MadGraph|. Different masses of the LQ fields are considered in this work, and for each parameter point we compute the total decay width in \verb|MadGraph|, which assumes the narrow-width approximation. Given that all possible decay modes may exist, depending on the coupling/mass ratio, the decay width may become larger than mass. In the numerical work that follows, we have only analysed points where the total decay width is always smaller than the mass. In practice, the signal process can be easily generated through the following set of \texttt{MadGraph} commands
	\begin{minted}{text}
	>> ./bin/mg5_aMC
	>> import model LQ_UFO_py3 
	>> define p = u u~ d d~ s s~ c c~ b b~ g 
	>> generate p p > mu- mu+ $$ a z h
	>> output LQ_T_channel
	>> launch
	\end{minted}
	Here, we note the 4th line (where we have added the string \texttt{\$\$ a z h} which eliminates the contributions from the SM part to the amplitude) neglects interference terms with the SM, which is a valid approximation for the mass scales that we consider in this work (above $1.5~\mathrm{TeV}$) \cite{Schmaltz:2018nls,Haisch:2022lkt}. In this regard, for some benchmark scenarios with masses between 1.5 and 3.5 TeV, we have estimated that the contributions from the interference terms amount to changes in the total cross section on the order of a few percent and therefore should not heavily impact the main results here. Indeed, as noted e.g. \cite{Azevedo:2022jnd}, interference terms that alter the total cross-section by around 10\% amounts to changes of a few events. Additionally, not all kinematic distributions may be sensitive to the interference terms and as such its effects may be hidden. However, is is important to note that this is not generic statement. Indeed, in some previous works \cite{Bhaskar:2022vgk,Bhaskar:2021pml,Aydemir:2019ynb,Mandal:2018kau} the interference between the $t-$channel and $\mathrm{Z^0}/\gamma$ SM contributions can play an important role in the exclusion limits, depending on the phase-space region (in particular, they can dominate the exclusion bounds for high values of the coupling/mass ratios). The most up-to-date versions of \texttt{MadGraph} are written in \texttt{python3} while the older versions are in \texttt{python2}. Therefore, we have made available both versions of the \texttt{UFO} files in the \texttt{GitHub} in \url{https://github.com/Mrazi09/LQ_collider_
		project} in the folders \texttt{LQ\_UFO\_py2} (for \texttt{python2}) and \texttt{LQ\_UFO\_py3} (for \texttt{python3}). For completeness, we would like to mention that we used version 3.4.1 of \texttt{MadGraph}. Additionally, \texttt{MadGraph} parameter space cards for the benchmark scenarios are also available in the \texttt{GitHub} page, in the folder named \texttt{Benchmark\_cards}. 
	
	
	We have generated a total of 10K events for the signal and 1M events for the main backgrounds ($t\bar{t}$, $\mathrm{Z^0}+\mathrm{jets}$ and $\mathrm{VV}+\mathrm{jets}$, where $V=\mathrm{Z^0}, W^\pm$). In all generated samples leptons were required to have transverse momenta $p_T \ge 25~\mathrm{GeV}$ and pseudo-rapidity $\eta \le 2.5$\footnote{The pseudo-rapidity is conventionally defined as $\eta = -\ln(\tan(\theta/2))$, where $\theta$ is the polar angle with respect to proton beam.}, at generator level. All generated collisions (backgrounds and signal) are simulated at a centre-of-mass (CM) energy of $\sqrt{s}=13.6~\mathrm{TeV}$, using the NNPDF2.3 parton distribution function \cite{Ball:2013hta}. We have fixed the top quark mass to $M_t = 173~\mathrm{GeV}$ and the $W$ ($Z$) boson mass to $M_W = 80.4~\mathrm{GeV}$($M_Z = 91.2~\mathrm{GeV}$). All other quarks are assumed to be massless. Muons are treated as stable particles with zero mass.
	\begin{figure*}[htb!]
		\centering
		\includegraphics[width=1.00\textwidth,height=\textheight,keepaspectratio]{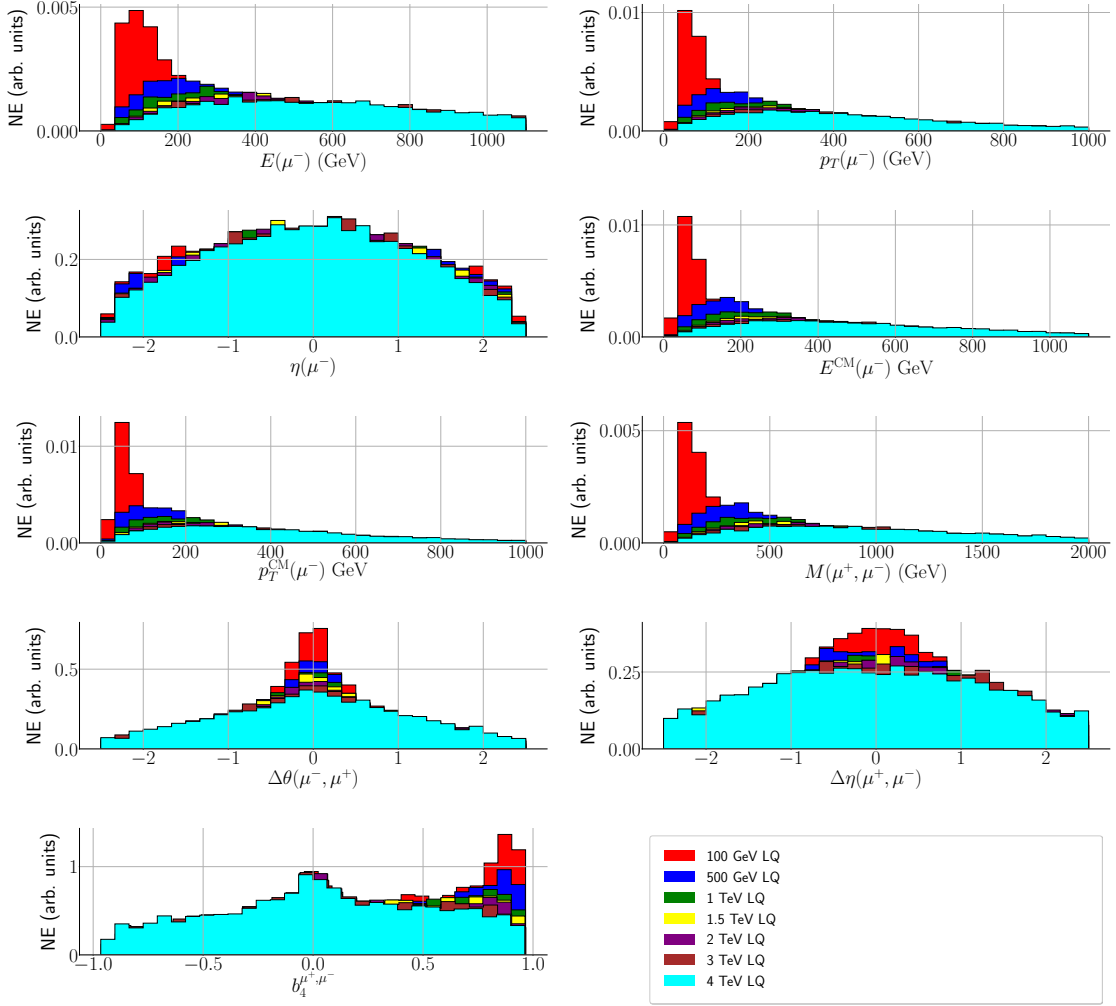}
		\vspace*{-4em}
		\caption{Normalised kinematic and angular distributions for various masses of the lightest LQ field with 30 bins each. The light LQ masses shown in the legend and are used merely to illustrate the mass dependence of the distributions. Here, $p_T$ is the transverse momentum, $E$ is the energy, $\eta$ is the pseudorapidity, $\theta$ is the polar angle and $M$ is the invariant mass. Equivalent distributions exist for the anti-muon.}
		\label{fig:LQ_observables_1}
	\end{figure*}
	
	It is interesting to notice that different mass scales affect kinematic distributions in a quite significant way. We show in Figure~\ref{fig:LQ_observables_1}, for different LQ masses (with the same couplings and mixings), the energy, the $p_T$ (both measured in CM and laboratory frames) and the pseudo-rapidity of the muons as well as the polar angle difference, the pseudo-rapidity difference and the invariant mass of the generated di-muon system. In addition, we also show the $b_4$ angular distribution, introduced in the context of $t\bar{t}h$ searches \cite{Ferroglia:2019qjy}, defined according to 
	\begin{equation}\label{eq:b4}
	b_4(i,j) = \frac{(p_{z,i}^f~p_{z,j}^f)}{\abs{\textbf{p}_i^f} \abs{\textbf{p}_j^f}} \, ,
	\end{equation}
	where $\textbf{p}^f$ is a three-momentum vector for $i,j = \mu^+,\mu^-$ (without any index overlap), $p_{z,i}^f$ is the total momentum along the $z$ direction. This observable is calculated only in the laboratory reference frame. 
	
	The muon's $p_T$ and energy, both in the CM and laboratory frame, as well as the dilepton invariant mass show an increase of events in the lower region whenever the mass is of order of 100 GeV, while long tails are expected in the LQs TeV mass range. Note that this behaviour also becomes apparent for certain angular distributions, in particular, the $\Delta \theta (\mu^+, \mu^-)$ and $\Delta \eta (\mu^+, \mu^-)$, where the distributions get longer tails with the mass increase. The muon $\eta$ distribution, on the contrary, does not show any particular structure dependence with the mass. The $b_4$ angular distribution is also of interest here, where we note that for lower masses, events tend to accumulate for higher values of the variable. Indeed, as we can observe, besides standard kinematic variables, angular distributions can act as an additional measure of the LQ mass, having the nice advantage of suffering less from experimental uncertainties compared to the other kinematic distributions.
	
	\subsection{Benchmark points}\label{subsec:benchmarks}
	
\begin{figure}[htb!]
		\centering
		\subfloat{\includegraphics[width=1.00\textwidth]{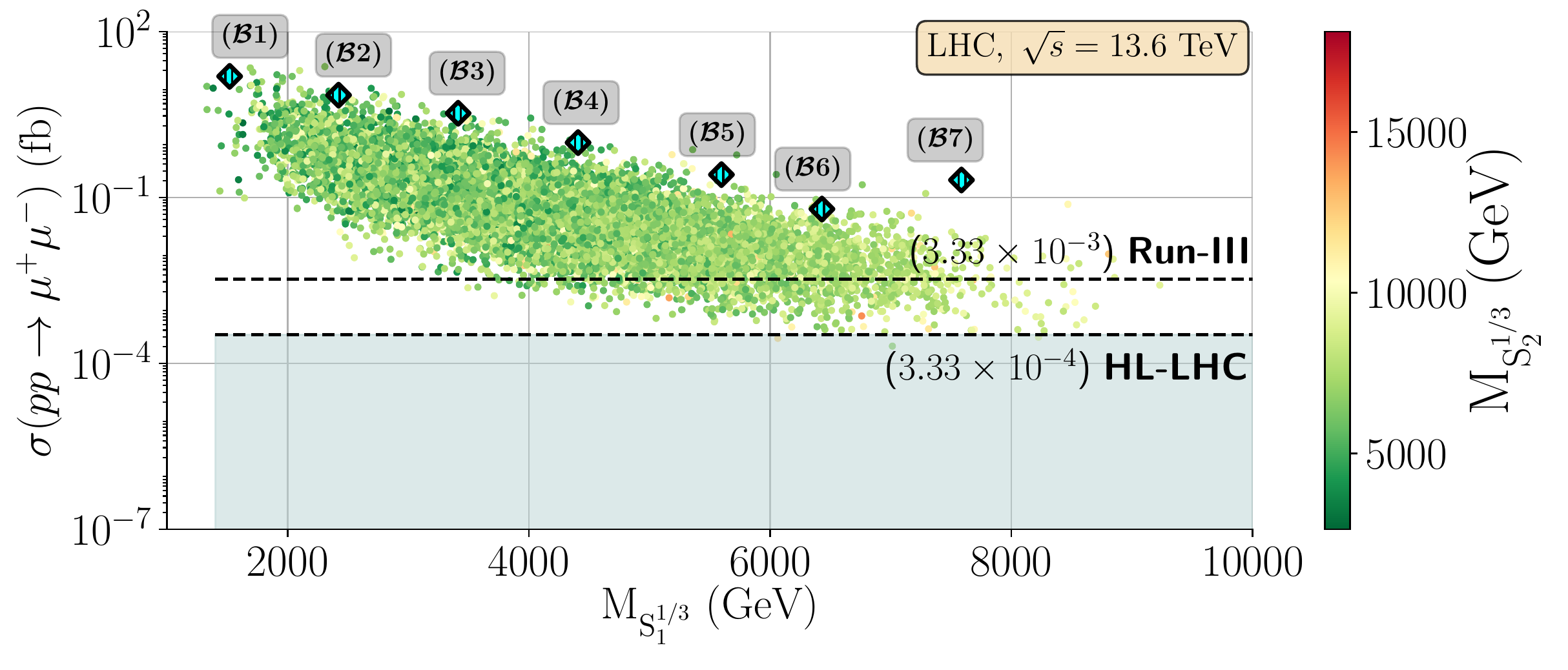}} 
		\caption{Total production cross-section ($\sigma$) in femtobarns (fb) as a function of the mass of the lightest LQ in GeV. In the colour axis we show the mass of the second-lightest LQ, also in GeV. Highlighted by horizontal dashed lines are represented the ones corresponding to at least one event at run-III and the HL-LHC phase, just to guide the eye. Please note that the total cross-section shown here is computed at generator level with \texttt{MadGraph}. Marked by cyan diamonds and an alphanumerical label are the benchmark points used in our analysis.}
		\label{fig:cross_section_plots}
	\end{figure}

	A comprehensive phenomenological analysis of our model \cite{Freitas:2022gqs}, taking into account up-to-date B-physics flavour observables ($R_{D,D^*}$, $R_{K,K^*}$), anomalous magnetic moment of the muon ($g-2$) and neutrino masses and mixings, was implemented. For each allowed point in the parameter space of the model, we have computed the total cross-section for the process $pp\rightarrow \mu^+\mu^-$. Results are shown in Fig.~\ref{fig:cross_section_plots}.
	
	We notice that a vast region of the parameter space can be probed within the predicted sensitivity ranges allowed for the future runs of the LHC (at both run-III, which is currently on-going, as well as at the high luminosity phase, HL-LHC). In particular, in Figure~\ref{fig:cross_section_plots}, we can see that masses of the lightest LQ $S_1^{1/3}$ of up to 8 TeV can be potentially probed already at run-III in the considered model. No estimate on backgrounds has been considered here, hence the limits shown here are merely qualitative and serve as a guideline for the potential mass scales where the LQ can be probed. In the colour axis (Fig.~\ref{fig:cross_section_plots}) we show the mass of the second-lightest $1/3e$ LQ. No evident correlation is found between this mass and the total cross-section which is primarily driven by the contribution of the lightest state. A similar conclusion can also be derived for the $2/3e$ LQ. This is directly related to the values considered for the coupling parameter ${\amber a_1}$ between the Higgs field and the LQs (see Eq.~\ref{eq:scalar_potential_2}) where we have considered a low value for the trilinear term (see appendix \ref{app:numerics}). This choice leads to a low mixing between the LQs (see Eq.~\ref{eq:mix_angle}). Hence, from this point onwards, when referring to the mass of the LQ, we always refer to the mass of the lightest LQ state.
	
	Based on these results, we have defined a set of benchmark points with different mass values of the lightest LQ state and cross-sections. The concrete values of couplings and masses used for the benchmark points are shown in Appendix~\ref{app:numerics}. We have computed, for these benchmark points, several kinematic and angular distributions for the final-state particles. 
	
	\section{Event selection}\label{sec:event_sel}
	
	A dedicated analysis was performed for each benchmark point. Following the generation and simulation by Delphes, all events were required to have at least two isolated charged muons with $p_T$ above 25 GeV and $\eta$ within the range of $[-2.5, 2.5]$. From the events that survive the selection criteria, we compute a variety of relevant kinematic/angular observables using the two highest $p_T$ muons. These include the transverse momenta, energy, pseudo-rapidity and the azimuthal angle of the muons. Additionally, we have considered the di-muon invariant mass $M(\mu^+,\mu^-)$, the cosine of the muon's polar angle difference $\cos \Delta \theta(\mu^-,\mu^+)$, the azimuthal angle difference $\Delta \Phi (\mu^+, \mu^-)$ and the muons $\Delta R (\mu^+, \mu^-)$ distribution~\footnote{$\Delta \theta(i,j)$ is defined as the difference between particles' $i$ and $j$ polar angles $\Delta \theta(i,j) = \theta_j - \theta_i$ and $\Delta R(i,j) = \sqrt{\Delta\Phi(i,j)^2 + \Delta\eta^2(i,j)}$, with $\Delta\Phi(i,j) = \Phi_j - \Phi_i$ and $\Delta \eta(i,j) = \eta_j - \eta_i$.}. The di-muon invariant mass distribution is reconstructed using $M^2 = ( p_{\mu^+} + p_{\mu^-})^2$, where $p_{\mu^\pm}$ are the muons four-momenta. Several of the observables are computed both in the laboratory reference frame and in the di-muon centre-of-mass frame.
	
	\section{Numerical results}\label{sec:results}
	\begin{figure*}[htb!]
		\centering
		\includegraphics[width=1.00\textwidth,height=\textheight,keepaspectratio]{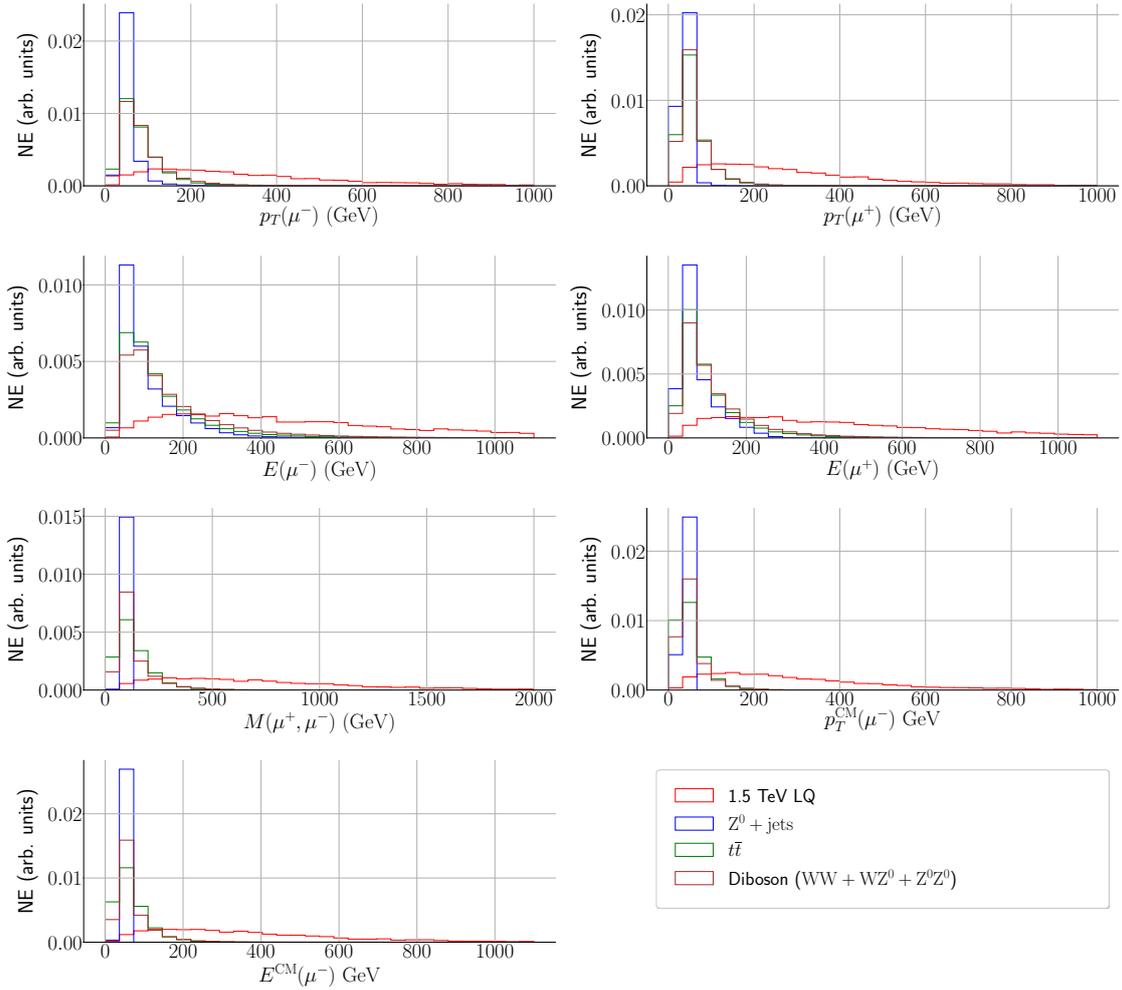}
		\caption{Normalised kinematic distributions for a LQ with 1.5 TeV mass (in red) and the main irreducible backgrounds ($\mathrm{Z^0+jets}$ in blue, $t\bar{t}$ in green and di-boson in brown) with 30 bins each. Here, $p_T$ is the transverse momentum, $E$ is the energy and $M$ is the invariant mass. Labels with a ``CM'' superscript are calculated in the muon/anti-muon centre-of-mass frame.}
		\label{fig:LQ_observables_2}
	\end{figure*}
	
	While it is interesting to look how the mass scale of LQ may impact the kinematical distributions, it is also of utmost relevance to understand the relative differences between the signal and the main irreducible backgrounds. We plot in Figure~\ref{fig:LQ_observables_2}, for the benchmark with a mass of $1.5~\mathrm{TeV}$, the most relevant distributions, i.e.~the ones that offer the greatest discriminating power for the signal with respect to the dominant backgrounds ($\mathrm{Z^0+}$jets, $t\bar{t}$ and di-boson). We note that the $p_T$, $E$ and $M(\mu^+,\mu^-)$  distributions offer the greatest signal discriminating power with respect to the SM backgrounds. The signal tends to populate the regions of higher values while the SM events are concentrated in the lower regions of these distributions. We have checked that for our signals, most of the angular distributions (not shown in Figure~\ref{fig:LQ_observables_2}) offered poor discriminant power, with the signal closely following the SM expectation and therefore we do not consider them hereafter. Taking into account the recommendations set forward by both ATLAS and CMS collaborations \cite{CERN_HLLHC, ATLAS:2022hro}, the estimated the systematic uncertainties are around 1\% - 1.5\%. Hence, we take the systematics to be 1\% for this analysis.
	
	The various kinematic distributions can then be used to estimate how the LQ di-muon signal (i.e.~the signal hypothesis $H_1$) can be discovered and exclude the SM hypothesis at the LHC (i.e.~the background/null hypothesis $H_0$) \cite{Cowan:2010js}, as a function of luminosity. We have simulated 50K pseudo-experiments based on Poisson fluctuations of $H_0$ and $H_1$. Two reference cases were considered i.e.~the full luminosity of run-III of the LHC, roughly around $300~\mathrm{fb^{-1}}$, and the full luminosity of the HL-LHC, at around $3000~\mathrm{fb^{-1}}$. The signal $p$-values, under the $H_0$ hypothesis, were extracted by calculating for each pseudo-experiment a test statistics, $\mathcal{F} (x_{i, \mathrm{S}};\, x_{i, \mathrm{N}})$, defined as the ratio between the probability of the pseudo-experiments being compatible with the signal hypothesis against the null hypothesis. We have computed the $p$-values for each benchmark point. 
	
\begin{figure*}[htb!]
		\centering
		\hspace*{-3.8em}
		\subfloat[LQ mass of $1.5~\mathrm{TeV}$ at $\mathcal{L}=300~\mathrm{fb^{-1}}$]{\includegraphics[width=0.62\textwidth]{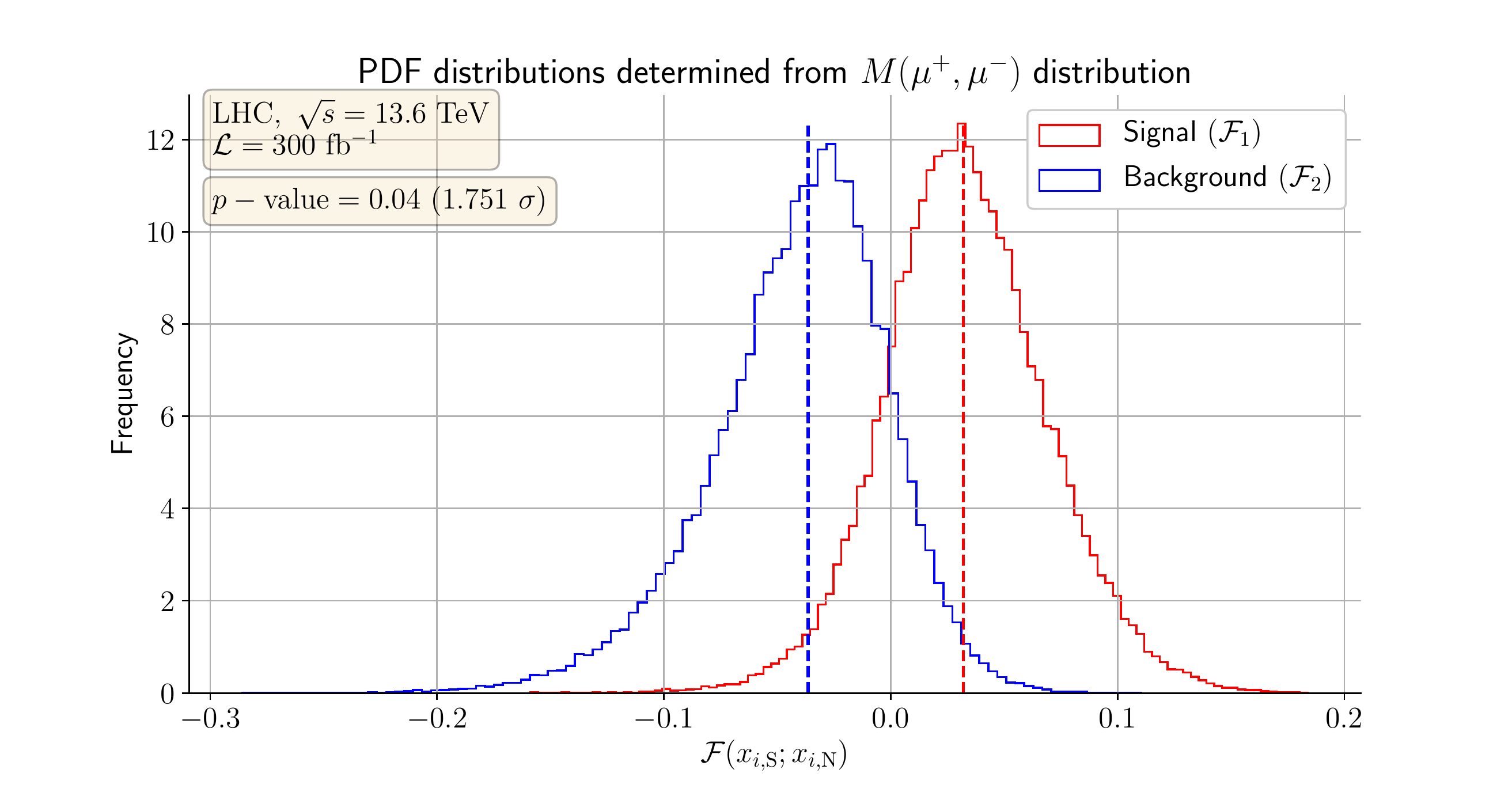}} \hspace*{-2.5em}
		\subfloat[LQ mass of $1.5~\mathrm{TeV}$ at $\mathcal{L}=3000~\mathrm{fb^{-1}}$]{\includegraphics[width=0.62\textwidth]{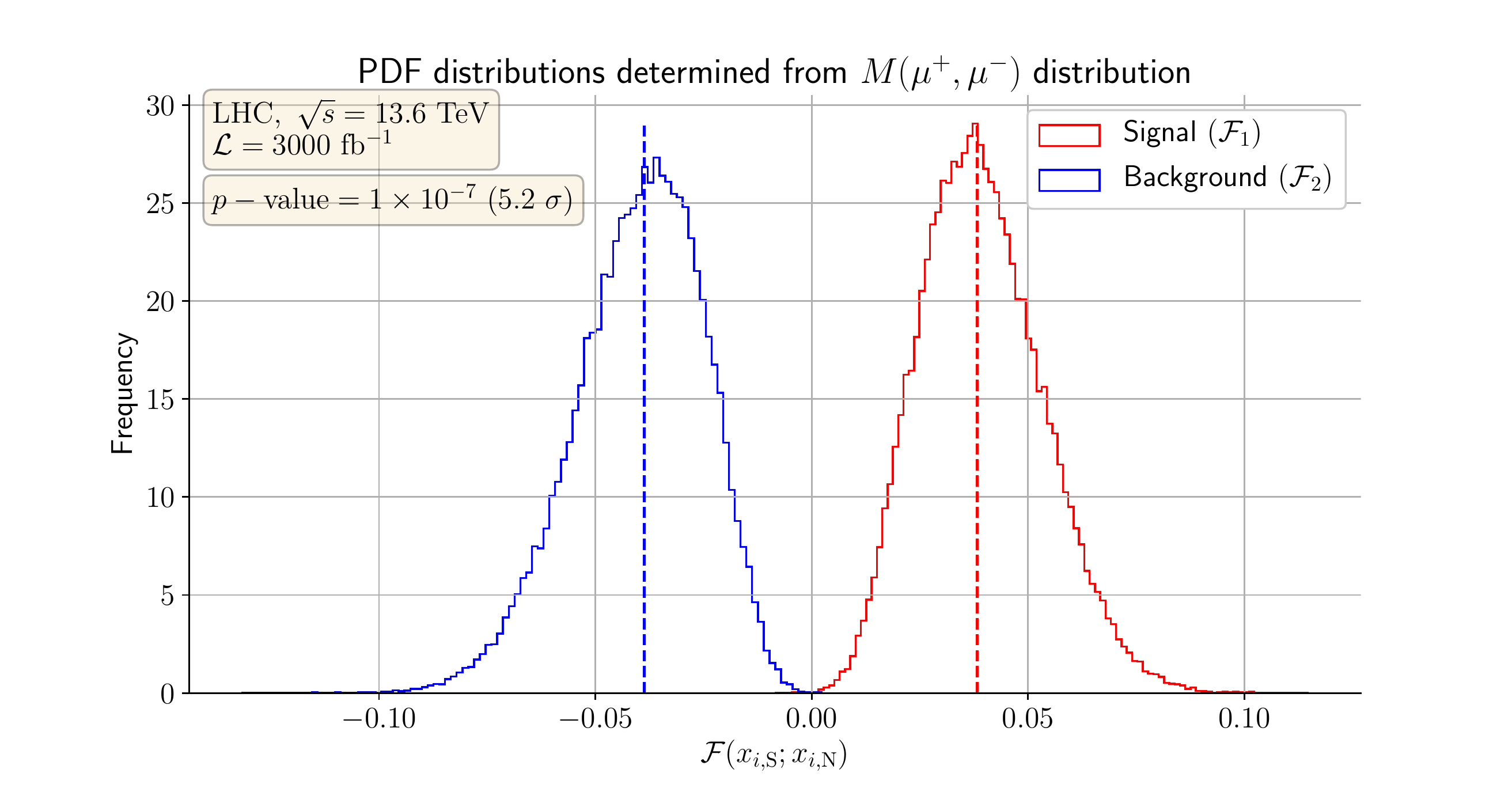}} 
		\caption{Histograms of frequency for the test statistic $\mathcal{F}(x_{i,\mathrm{S}}; x_{i,\mathrm{N}})$ based on the $M(\mu^+,\mu^-)$ distribution for the benchmark with the mass of the LQ as 1.5 TeV.}
		\label{fig:pt_gaussfit}
	\end{figure*}

	We show in Figure~\ref{fig:pt_gaussfit} examples of the test statistics $\mathcal{F} (x_{i, \mathrm{S}};\, x_{i, \mathrm{N}})$ for the 1.5 TeV benchmark point, assuming the run-III luminosity (left) and the HL-LHC luminosity (right), using the di-muon invariant mass. Both $H_1$ and $H_0$ are shown. We notice that, by just using this distribution, a small fluctuation of $1.75\sigma$ in $H_0$ at run-III is sufficient to achieve the signal level i.e.~effectively mimicking the signal by the background fluctuation. At the HL-LHC, a $5.20\sigma$ fluctuation in $H_0$ would be required to cross the discovery threshold. Of course, the di-muon invariant mass is not the only distribution we can use, as the $p_T$ and energy distributions can also give potentially equivalent results (see Figure~\ref{fig:LQ_observables_2}). For completeness, we show in Table~\ref{tab:significance_table} all results for the remainder benchmarks scenarios, where we see that the combination of the various distributions can now extend the discovery threshold of our model's LQs to larger mass scales, that could go up to roughly 2.5 TeV, at the HL-LHC. 
	{\renewcommand{\arraystretch}{0.9}%
		\begin{table}[htb!]
			\begin{center}
				\captionsetup{justification=raggedright,singlelinecheck=true}
				\resizebox{\columnwidth}{!}{%
					\begin{tabular}{|c|c|c|c|c|c|}\hline
						$M_{LQ}~(\mathrm{GeV})$ & \multicolumn{1}{c|}{\makecell{$M(\mu^+,\mu^-)$ \\ $\mathcal{L} = 300~\mathrm{fb^{-1}}$}} & \multicolumn{1}{c|}{\makecell{$M(\mu^+,\mu^-)$ \\ $\mathcal{L} = 3000~\mathrm{fb^{-1}}$}} & \multicolumn{1}{c|}{\makecell{$E(\mu^+)$ \\ $\mathcal{L} = 300~\mathrm{fb^{-1}}$}} & \multicolumn{1}{c|}{\makecell{$E(\mu^+)$ \\ $\mathcal{L} = 3000~\mathrm{fb^{-1}}$}} & \multicolumn{1}{c|}{\makecell{$\mathrm{Combined}$ \\ $(300,3000)~\mathrm{fb^{-1}}$}} \\[2mm] \hline
						\multirow{2}{*}{$1.5~\mathrm{TeV}~(\mathcal{B}1)$} & \multirow{2}{*}{$1.75\sigma$} & \multirow{2}{*}{{\redBU $5.20\sigma$}} & \multirow{2}{*}{$0.891\sigma$} & \multirow{2}{*}{$2.72\sigma$} & \multirow{2}{*}{($3.06\sigma$,\, {\redBU $9.72\sigma$})} \\[5mm] \hline
						\multirow{2}{*}{$2.5~\mathrm{TeV}~(\mathcal{B}2)$} & \multirow{2}{*}{$0.573\sigma$} & \multirow{2}{*}{$2.28\sigma$} & \multirow{2}{*}{$0.744\sigma$} & \multirow{2}{*}{$2.33\sigma$} & \multirow{2}{*}{($1.15\sigma$,\, $4.97\sigma$)} \\[5mm] \hline
						\multirow{2}{*}{$3.5~\mathrm{TeV}~(\mathcal{B}3)$} & \multirow{2}{*}{$0.128\sigma$} & \multirow{2}{*}{$0.912\sigma$} & \multirow{2}{*}{$0.225\sigma$} & \multirow{2}{*}{$1.04\sigma$} & \multirow{2}{*}{($0.288\sigma$,\, $1.97\sigma$)} \\[5mm] \hline
						\multirow{2}{*}{$4.5~\mathrm{TeV}~(\mathcal{B}4)$} & \multirow{2}{*}{$0.000698\sigma$} & \multirow{2}{*}{$0.181\sigma$} & \multirow{2}{*}{$0.0072\sigma$} & \multirow{2}{*}{$0.360\sigma$} & \multirow{2}{*}{($0.0140\sigma$,\, $0.356\sigma$)} \\[5mm] \hline
						\multirow{2}{*}{$5.5~\mathrm{TeV}~(\mathcal{B}5)$} & \multirow{2}{*}{$0.00154\sigma$} & \multirow{2}{*}{$0.0161\sigma$} & \multirow{2}{*}{$0.000421\sigma$} & \multirow{2}{*}{$0.124\sigma$} & \multirow{2}{*}{($0.0139\sigma$,\, $0.05\sigma$)} \\[5mm] \hline
						\multirow{2}{*}{$6.5~\mathrm{TeV}~(\mathcal{B}6)$} & \multirow{2}{*}{$0.00419\sigma$} & \multirow{2}{*}{$0.0272\sigma$} & \multirow{2}{*}{$0.0112\sigma$} & \multirow{2}{*}{$0.00958\sigma$} & \multirow{2}{*}{($0.00339\sigma$,\, $0.0238\sigma$)} \\[5mm] \hline
						\multirow{2}{*}{$7.5~\mathrm{TeV}~(\mathcal{B}7)$} & \multirow{2}{*}{$0.00789\sigma$} & \multirow{2}{*}{$0.0292\sigma$} & \multirow{2}{*}{$0.0027\sigma$} & \multirow{2}{*}{$0.0635\sigma$} & \multirow{2}{*}{($0.00319\sigma$,\, $0.0509\sigma$)} \\[5mm] \hline
				\end{tabular}}
				\caption{The statistical significance for each of the mass benchmarks mentioned in the text for $M(\mu^+,\mu^-)$ and $p_T(\mu^+)$ distributions. In {\redBU red} we indicate the point that passes the $5\sigma$ threshold for discovery. In the last column we show the combined significance, calculated based on the distributions shown in Fig.~\ref{fig:LQ_observables_2}.}
				\label{tab:significance_table}
			\end{center}
	\end{table}}
	
\begin{figure}[htb!]
		\centering
		\subfloat{\includegraphics[width=1.00\textwidth]{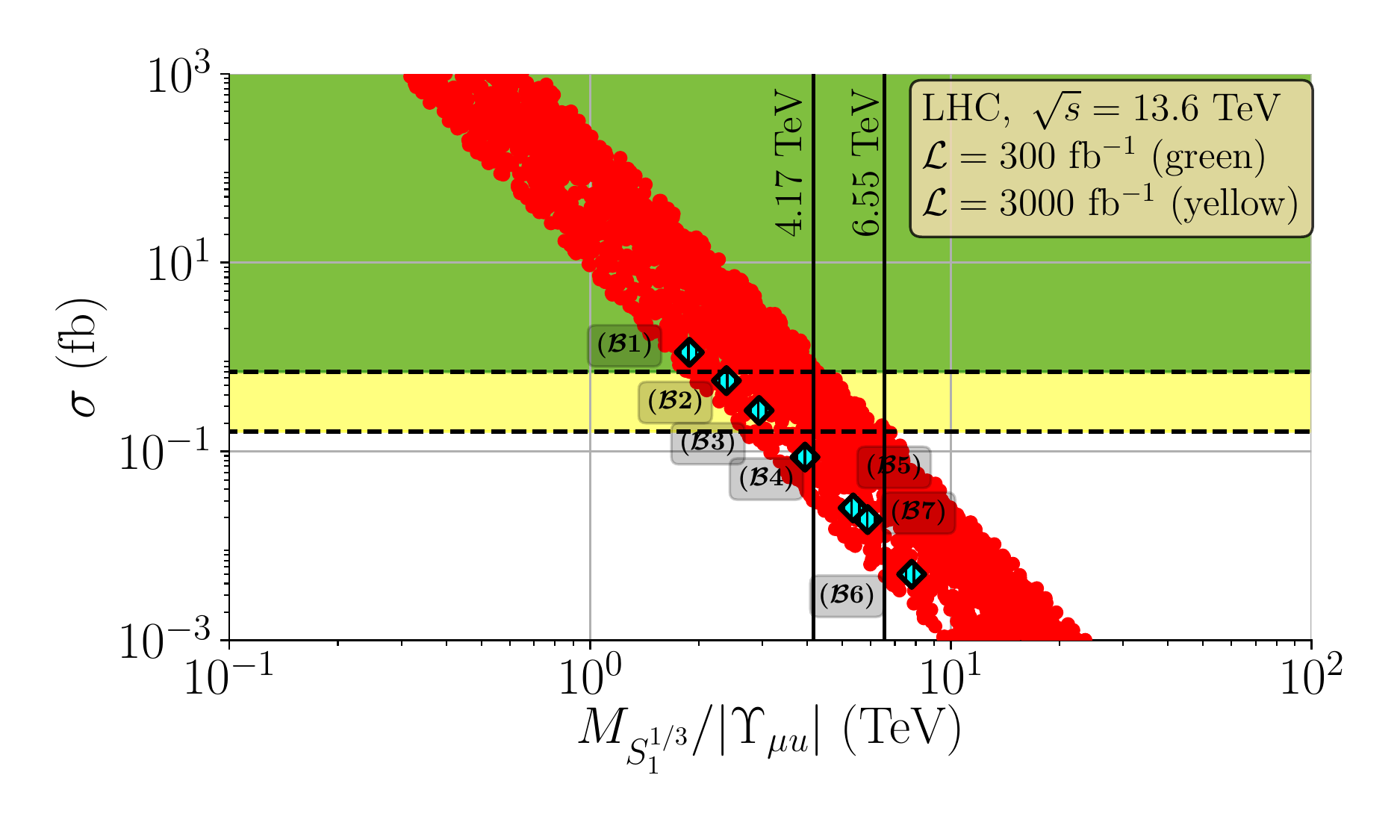}} 
		\caption{Total cross-section, in fb, as a function of the coupling/mass ratio in TeV. The area marked in green represents the region where the point can be excluded at a 95\% CL during run-III, whereas the yellow region is the equivalent for the HL-LHC. Both the y-axis and the x-axis are in logarithmic scale. Benchmark points are marked by cyan diamonds.}
		\label{fig:xsec_MUpsilon}
	\end{figure}

	The previous results show great promise in excluding the model in the future runs of the LHC, although, they do not offer a complete picture. In fact, we are analysing a $t$-channel mediated process where the coupling between the LQ field and the quarks that originate from the proton plays a vital role. To this end, we show in Figure~\ref{fig:cross_section_plots} a scatter plot of the total cross section in fb as a function of the mass/coupling ratio in TeV. We have found that the main coupling that mediates the process in Figure~\ref{fig:feyn_diagrams} is ${\color{ForestGreen}\Upsilon_{\mu u}}$\footnote{Other couplings can also mediate this process. However, looking at the numerical values of the couplings that are in Appendix~\ref{app:numerics} shows that the remainder couplings are much smaller. A thorough study on the implications of the remainder entries is beyond the scope of this paper and left for future explorations.}. To determine the bounds coming from run-III (HL-LHC) shown in green (yellow) in Figure~\ref{fig:xsec_MUpsilon}, we have picked the $1.5~\mathrm{TeV}$ benchmark point and re-scaled the total cross-section to find the minimum value of the cross-section ($\sigma_{\mathrm{min}}$) that still allows for exclusion of $H_0$ at 95\% CL. For run-III we obtained $\sigma_{\mathrm{min}} \sim 0.70~\mathrm{fb}$ and for HL-LHC $\sigma_{\mathrm{min}} \sim 0.16~\mathrm{fb}$. The points shown in Figure~\ref{fig:xsec_MUpsilon} are determined by varying only the ${\color{ForestGreen}\Upsilon_{\mu u}}$ coupling and the mass of the lightest LQ field, with all other coupling parameters fixed to the same values of benchmark $\mathcal{B}1$ (see Appendix \ref{app:numerics}). To avoid interference effects between the different LQ states, besides imposing a low mixing between the fields with a small ${\amber a_1}$, the masses of the remainder fields were made arbitrary large, around $50~\mathrm{TeV}$, such that only the lightest field can impact the cross-section. Additionally, for each shown point, the decay width is automatically computed through \texttt{MadGraph}.
	
	From Figure~\ref{fig:xsec_MUpsilon} we can clearly see a correlation in logarithmic scale between the cross-section of the process and the ratio between the mass of the lightest LQ and the ${\color{ForestGreen} \Upsilon_{\mu u}}$. This helps to demonstrate the relevance of the couplings in the process, complementing nicely the results shown before in Table~\ref{tab:significance_table}. Indeed, the benchmark points of $1.5,\, 2.5~\mathrm{and}~3.5~\mathrm{TeV}$ which correspond to the ones that we are able to exclude at $\sim$95\% CL at the HL-LHC, are represented by the first three points in the plot. Only one point appears in the green region, for instance the $1.5~\mathrm{TeV}$ benchmark. In particular, the mass/coupling ratio of these points reads as $m_{S^{1/3}_1}/\abs{\color{ForestGreen} \Upsilon_{\mu u}} = 1.885~\mathrm{TeV}$, $m_{S^{1/3}_1}/\abs{\color{ForestGreen} \Upsilon_{\mu u}} = 2.389~\mathrm{TeV}$ and $m_{S^{1/3}_1}/\abs{\color{ForestGreen} \Upsilon_{\mu u}} = 2.942~\mathrm{TeV}$. Based on these results, at 95\% CL we have determined that at run-III one can exclude the model's mass/coupling ratios up to $4.17~\mathrm{TeV}$, which can be further improved to $6.55~\mathrm{TeV}$ if one considers the HL-LHC. 
	
	\section{Conclusions}\label{sec:conclusions}
	
	In this work, we have performed a comprehensive collider analysis of a minimal LQ model composed of a single generation of a weak-singlet state and a weak-doublet state, leading to three new physical scalar particles. As it was shown in a previous work by some of the authors \cite{Freitas:2022gqs}, this framework provides a radiative mechanism for the generation of neutrino masses and mixing and is consistent with the most stringent flavour observables. It can also explain the $W$-mass anomaly, the anomalous magnetic moment of the muon and the $R_{D,D^*}$ anomalies if experiments confirm such deviations. We have performed the collider phenomenology analysis for a series of benchmark points, focused on a di-muon final state, which is mediated by the scalar LQs in the $t$-channel. We have shown that there is plenty of parameter space domains available for the model to be probed at planned runs of the LHC. In particular, we have found that at run-III one can exclude mass/coupling ratios of up to $4.17~\mathrm{TeV}$, while for the HL-LHC phase, the exclusion limits become more stringent, hitting an upper bound of $6.55~\mathrm{TeV}$ at 95\% CL. 
	
	\begin{figure*}[htb!]
		\centering
		\subfloat{\includegraphics[width=0.40\textwidth]{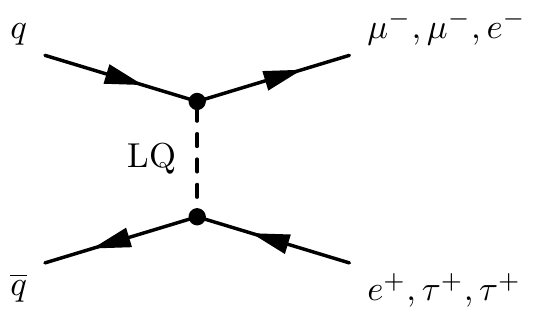}}
		\hspace*{3em}
		\subfloat{\includegraphics[width=0.32\textwidth]{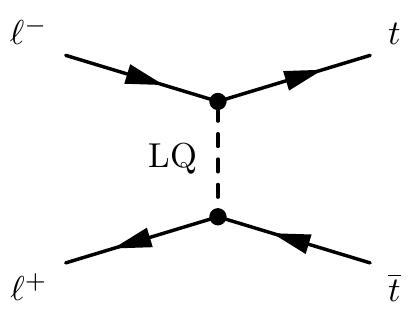}} 
		\caption{Topologies for non-ressonant production of scalar LQs for flavour off-diagonal final states (on the left) and for top-quark pair production (on the right).}
		\label{fig:top_concl}
	\end{figure*}
	
	Although we have focused on the simplest scenario for the topology in a muon-pair final state, various other combinations are possible for the final state particles, such as the $e^+e^-$ and $\tau^+\tau^-$ channels, which can provide further constraints for the LQ couplings with the first and third generations, which as noted in the introduction, represent couplings with the weaker constraints. Additionally, since a non-diagonal structure is present in the Yukawa couplings, more exotic channels such as e.g. $q\bar{q}\rightarrow \mu^-\tau^+$ (Figure \ref{fig:top_concl}) can be probed provided the couplings ${\color{ForestGreen} \Upsilon_{\mu q}}$ and ${\color{ForestGreen} \Upsilon_{\tau q}}$ are strong enough to be visible at the LHC. As we have noted before, we have found ${\color{ForestGreen} \Upsilon}$ to be the main driving force for the $t$-channel process, although, both ${\color{red} \Theta}$ and ${\color{blue} \Omega}$ can also mediate this topology (in essence, it would simply correspond to replacing the up-quark with a down-quark in the initial proton). Although these couplings connect to the down sector, mediating certain processes such as Kaon decays and meson mixing, which are tightly constrained by experiments, their sizes are expected to be small. Indeed, this was the conclusion found in \cite{Freitas:2022gqs}. On the other hand, ${\color{red} \Theta}$ and ${\color{blue} \Omega}$ couplings involving third/second generation fermions are not as tightly constrained. Indeed, based on the textures shown in \cite{Freitas:2022gqs}, channels that involve the top-muon-LQ coupling can also be interesting to explore in future works, in particular in the context of future planned lepton colliders, where top quark pairs can be pair-produced in the $t$-channel process (see the diagram on the left of Fig.~\ref{fig:top_concl}).
	
	\acknowledgments
	The authors would like to thank Werner Porod for insightful discussions that contributed to the final shape of this manuscript.
	The authors would also like to thank Tanumoy Mandal for pointing out the impact of the interference between the SM and LQ diagrams and Alexander Belyaev for discussions on the impact of interference effects and systematic errors on the exclusion bounds.
	J.G. and A.P.M.~are supported by the Center for Research and Development in Mathematics and Applications (CIDMA) through the Portuguese Foundation for Science and Technology (FCT - Funda\c{c}\~{a}o para a Ci\^{e}ncia e a Tecnologia), references UIDB/04106/2020 and UIDP/04106/2020. A.P.M. is supported by the project PTDC/FIS-PAR/31000/2017. A.P.M. and  J.G. are also supported by the projects CERN/FIS-PAR/0019/2021, CERN/FIS-PAR/0021/2021, CERN/FIS-PAR/0019/2021 and CERN/FIS-PAR/0025/2021
	J.G. is also directly funded by FCT through the doctoral program grant with the reference 2021.04527.BD.
	A.P.M.~is also supported by national funds (OE), through FCT, I.P., in the scope of the framework contract foreseen in the numbers 4, 5 and 6 of the article 23, of the Decree-Law 57/2016, of August 29, changed by Law 57/2017, of July 19.
	R.P.~and J.G.~are supported in part by the Swedish Research Council grant, contract number 2016-05996, as well as by the European Research Council (ERC) under the European Union's Horizon 2020 research and innovation programme (grant agreement No 668679).
	A.O. acknowledges support from national funds from FCT, under the project CERN/FIS-PAR/0037/2021.
	
	\appendix
	\section{Numerical values for relevant parameters}\label{app:numerics}
	
	In this section, we write the relevant numerical parameters for the various benchmarks used in this work. They correspond to the cyan diamonds shown in Figure~\ref{fig:cross_section_plots}. The numerical values are shown in the same order as in the plot, i.e.~they are ordered by mass. We also stress that \texttt{MadGraph} parameter cards for these benchmarks are already available in the \texttt{GitHub} page, as well as both \texttt{python2} and \texttt{python3} UFO files.
	\\ \\
	\underline{Benchmark 1 ($\mathcal{B}1$):}
	
	\begin{equation*}\nonumber
	{\color{ForestGreen} \Upsilon} = 
	\begin{pmatrix}
	-9.95 \times 10^{-7}+3.2 \times 10^{-7}i & 0.000614-0.001498i & -2.352 \times 10^{-7}-1.623 \times 10^{-7}i \\
	0.7929+0.1443i & -1.867 \times 10^{-8}+6.755 \times 10^{-8}i & -0.57+1.501i \\
	-2.184 \times 10^{-8}-4.347 \times 10^{-8}i & -0.00735-0.0115i & 0.0007205+0.0006524i
	\end{pmatrix}
	\end{equation*}
	
	\begin{equation*}\nonumber
	{\color{red} \Theta} = 
	\begin{pmatrix}
	-0.000853-0.003701i & -3.397 \times 10^{-7}+4.33 \times 10^{-8}i & 3.17 \times 10^{-8}-6.358 \times 10^{-7}i \\
	-0.0002189+0.0003971i & -6.815 \times 10^{-7}-5.32 \times 10^{-8}i & 1.378+0.244i \\
	0.03014-0.03425i & 0.005574-0.006047i & 0.479+0.952i
	\end{pmatrix}
	\end{equation*}
	
	\begin{equation*}\nonumber
	{\color{blue} \Omega} = 
	\begin{pmatrix}
	-0.00379+0.02799i & 0.06373-0.03427i & -0.02809-0.0868i \\
	8.424 \times 10^{-7}+1.55 \times 10^{-8}i & 6.207 \times 10^{-6}-7.7 \times 10^{-8}i & -5.54 \times 10^{-7}-1.237 \times 10^{-6}i \\
	0.01112+0.02178i & -0.05615+0.06251i & -0.0059-0.1276i
	\end{pmatrix}
	\end{equation*}
	
	\begin{equation*}\nonumber
	m_{S^{1/3}_1} = 1519.208~\mathrm{GeV},\,\quad m_{S^{1/3}_2} = 4908.943~\mathrm{GeV},\,\quad m_{S^{2/3}} = 4900.669~\mathrm{GeV}
	\end{equation*}
	
	\begin{equation*}\nonumber
	{\amber a_1} = 3.31~\mathrm{GeV},\,\quad \sin 2\theta = -2.15 \times 10^{-7}
	\end{equation*}
	\\ 
	\underline{Benchmark 2 ($\mathcal{B}2$):}
	
	\begin{equation*}\nonumber
	{\color{ForestGreen} \Upsilon} = 
	\begin{pmatrix}
	-6.39 \times 10^{-7}+1.028 \times 10^{-6}i & 0.002902-0.003127i & -3.909 \times 10^{-8}-4.696 \times 10^{-8}i \\
	0.311+0.965i & -4.56 \times 10^{-8}+1.351 \times 10^{-7}i & -0.0108-0.0266i \\
	-4.811 \times 10^{-7}-2.01 \times 10^{-7}i & -0.679+1.757i & 0.000952+0.000428i
	\end{pmatrix}
	\end{equation*}
	
	\begin{equation*}\nonumber
	{\color{red} \Theta} = 
	\begin{pmatrix}
	0.004525-0.00427i & 4.361 \times 10^{-7}+5.852 \times 10^{-7}i & -8.553 \times 10^{-8}-2.118 \times 10^{-8}i \\
	-0.0002189+0.0001213i & 8.156 \times 10^{-7}-1.67 \times 10^{-7}i & 0.8792+0.0779i \\
	0.03981+0.02048i & 0.00526+0.01232i & 0.4013+0.7769i
	\end{pmatrix}
	\end{equation*}
	
	\begin{equation*}\nonumber
	{\color{blue} \Omega} = 
	\begin{pmatrix}
	0.006284+0.007269i & 0.00464+0.01796i & -0.03112-0.04218i \\
	6.982 \times 10^{-7}+3.19 \times 10^{-8}i & 8.687 \times 10^{-6}-6.5 \times 10^{-8}i & -3.689 \times 10^{-7}-7.436 \times 10^{-7}i \\
	0.000567+0.005244i & 0.00997-0.00793i & -0.00752-0.02016i
	\end{pmatrix}
	\end{equation*}
	
	\begin{equation*}\nonumber
	m_{S^{1/3}_1} = 2421.978~\mathrm{GeV},\,\quad m_{S^{1/3}_2} = 4177.546~\mathrm{GeV},\,\quad m_{S^{2/3}} = 4162.258~\mathrm{GeV}
	\end{equation*}
	
	\begin{equation*}\nonumber
	{\amber a_1} = 8.09~\mathrm{GeV},\,\quad \sin 2\theta = -9.881 \times 10^{-7}
	\end{equation*}
	\\
	\underline{Benchmark 3 ($\mathcal{B}3$):}
	
	\begin{equation*}\nonumber
	{\color{ForestGreen} \Upsilon} = 
	\begin{pmatrix}
	-1.058 \times 10^{-6}+7.87 \times 10^{-7}i & 0.000916-0.00437i & -2.149 \times 10^{-8}-2.316 \times 10^{-8}i \\
	1.128+0.27i & -2.179 \times 10^{-8}+4.237 \times 10^{-8}i & -0.00492-0.00893i \\
	-2.361 \times 10^{-7}-1.424 \times 10^{-7}i & -0.648+1.56i & 0.0007151+0.0003861i
	\end{pmatrix}
	\end{equation*}
	
	\begin{equation*}\nonumber
	{\color{red} \Theta} = 
	\begin{pmatrix}
	-0.00399-0.01319i & 1.261 \times 10^{-7}+3.1 \times 10^{-8}i & -1.835 \times 10^{-7}+3.101 \times 10^{-7}i \\
	-0.0001348+0.0003201i & -2.994 \times 10^{-8}-2.924 \times 10^{-8}i & 1.625+0.122i \\
	0.02901-0.03383i & 0.005963-0.002257i & 0.3007+0.6667i
	\end{pmatrix}
	\end{equation*}
	
	\begin{equation*}\nonumber
	{\color{blue} \Omega} = 
	\begin{pmatrix}
	-0.00835+0.01383i & 0.05926+0.01807i & -0.03707-0.06703i \\
	1.461 \times 10^{-6}+1.2 \times 10^{-8}i & 4.536 \times 10^{-6}-5.2 \times 10^{-8}i & -6.7 \times 10^{-7}-6.81 \times 10^{-7}i \\
	-0.00862+0.0066i & 0.1097+0.0959i & -0.0919-0.1284i
	\end{pmatrix}
	\end{equation*}
	
	\begin{equation*}\nonumber
	m_{S^{1/3}_1} = 3413.435~\mathrm{GeV},\,\quad m_{S^{1/3}_2} = 8300.163~\mathrm{GeV},\,\quad m_{S^{2/3}} = 8281.753~\mathrm{GeV}
	\end{equation*}
	
	\begin{equation*}\nonumber
	{\amber a_1} = 7.82~\mathrm{GeV},\,\quad \sin 2\theta = -1.933 \times 10^{-7}
	\end{equation*}
	\\
	\underline{Benchmark 4 ($\mathcal{B}4$):}
	
	\begin{equation*}\nonumber
	{\color{ForestGreen} \Upsilon} = 
	\begin{pmatrix}
	-8.15 \times 10^{-7}+7.12 \times 10^{-7}i & 0.000722-0.001487i & -1.203 \times 10^{-8}-8.081 \times 10^{-8}i \\
	1.035+0.421i & -4.547 \times 10^{-8}+8.488 \times 10^{-8}i & -0.00586-0.00922i \\
	-1.656 \times 10^{-7}-9.23 \times 10^{-8}i & -0.432+0.925i & 0.0006271+0.0003712i
	\end{pmatrix}
	\end{equation*}
	
	\begin{equation*}\nonumber
	{\color{red} \Theta} = 
	\begin{pmatrix}
	0.00668-0.01107i & 7.922 \times 10^{-7}+6.24 \times 10^{-8}i & 3.663 \times 10^{-7}-3.732 \times 10^{-7}i \\
	-0.0001807+0.0001545i & 8.592 \times 10^{-7}+1.43 \times 10^{-8}i & 2.01+0.232i \\
	0.06753+0.0513i & 0.00782+0.03077i & 0.922+0.714i
	\end{pmatrix}
	\end{equation*}
	
	\begin{equation*}\nonumber
	{\color{blue} \Omega} = 
	\begin{pmatrix}
	0.00795+0.01398i & -0.00649+0.018i & -0.02646-0.06841i \\
	6.931 \times 10^{-7}+2 \times 10^{-8}i & 9.014 \times 10^{-6}-7.3 \times 10^{-8}i & -3.08 \times 10^{-7}-1.076 \times 10^{-6}i \\
	-0.00172-0.001926i & 0.02823-0.0709i & -0.01426+0.06033i
	\end{pmatrix}
	\end{equation*}
	
	\begin{equation*}\nonumber
	m_{S^{1/3}_1} = 4407.591~\mathrm{GeV},\,\quad m_{S^{1/3}_2} = 5094.504~\mathrm{GeV},\,\quad m_{S^{2/3}} = 5075.82~\mathrm{GeV}
	\end{equation*}
	
	\begin{equation*}\nonumber
	{\amber a_1} = 4.94~\mathrm{GeV},\,\quad \sin 2\theta = -1.069 \times 10^{-6}
	\end{equation*}
	\\
	\underline{Benchmark 5 ($\mathcal{B}5$):}
	
	\begin{equation*}\nonumber
	{\color{ForestGreen} \Upsilon} = 
	\begin{pmatrix}
	-4.511 \times 10^{-7}+3.354 \times 10^{-7}i & 0.001305-0.003072i & -2.535 \times 10^{-8}-3.615 \times 10^{-8}i \\
	0.701+0.773i & -4.208 \times 10^{-8}+4.38 \times 10^{-8}i & -0.01668-0.02451i \\
	-1.173 \times 10^{-7}-1.852 \times 10^{-7}i & -0.303+2.223i & 0.0004531+0.0002779i
	\end{pmatrix}
	\end{equation*}
	
	\begin{equation*}\nonumber
	{\color{red} \Theta} = 
	\begin{pmatrix}
	-0.01321-0.02681i & 4.763 \times 10^{-8}-1.069 \times 10^{-8}i & -9.217 \times 10^{-7}-1.35 \times 10^{-8}i \\
	-0.000302+0.0001654i & 1.246 \times 10^{-8}-1.172 \times 10^{-8}i & 2.178+0.198i \\
	0.04608-0.07892i & 0.003842-0.005358i & 0.3288+0.9393i
	\end{pmatrix}
	\end{equation*}
	
	\begin{equation*}\nonumber
	{\color{blue} \Omega} = 
	\begin{pmatrix}
	-0.001847+0.00456i & 0.04349+0.01124i & -0.03003-0.03459i \\
	5.535 \times 10^{-7}+4.61 \times 10^{-8}i & 8.044 \times 10^{-6}-1.22 \times 10^{-7}i & -2.92 \times 10^{-7}-1.048 \times 10^{-6}i \\
	0.001107+0.005808i & -0.0278+0.03445i & 0.01955-0.06251i
	\end{pmatrix}
	\end{equation*}
	
	\begin{equation*}\nonumber
	m_{S^{1/3}_1} = 5596.399~\mathrm{GeV},\,\quad m_{S^{1/3}_2} = 7366.804~\mathrm{GeV},\,\quad m_{S^{2/3}} = 7392.307~\mathrm{GeV}
	\end{equation*}
	
	\begin{equation*}\nonumber
	{\amber a_1} = 10.62~\mathrm{GeV},\,\quad \sin 2\theta = -6.543 \times 10^{-7}
	\end{equation*}
	\\
	\underline{Benchmark 6 ($\mathcal{B}6$):}
	
	\begin{equation*}\nonumber
	{\color{ForestGreen} \Upsilon} = 
	\begin{pmatrix}
	-4.32 \times 10^{-7}+1.074 \times 10^{-6}i & 0.001191-0.00246i & -1.821 \times 10^{-8}-2.013 \times 10^{-8}i \\
	0.7528+0.3372i & -1.812 \times 10^{-8}+6.559 \times 10^{-8}i & -0.00851-0.01993i \\
	-2.195 \times 10^{-7}-1.287 \times 10^{-7}i & -0.347+3.145i & 0.0005461+0.0004567i
	\end{pmatrix}
	\end{equation*}
	
	\begin{equation*}\nonumber
	{\color{red} \Theta} = 
	\begin{pmatrix}
	0.01304+0.00986i & -1.455 \times 10^{-8}+4.789 \times 10^{-8}i & -7.5 \times 10^{-8}+3.563 \times 10^{-7}i \\
	-0.0001443+0.0002143i & 9.51 \times 10^{-7}+4.31 \times 10^{-7}i & 2.54+0.341i \\
	0.01175-0.02652i & -0.0032-0.0095i & 0.399+1.01i
	\end{pmatrix}
	\end{equation*}
	
	\begin{equation*}\nonumber
	{\color{blue} \Omega} = 
	\begin{pmatrix}
	0.00814-0.0076i & -0.0089+0.1034i & -0.02671-0.06354i \\
	6.259 \times 10^{-7}+2.32 \times 10^{-8}i & 5.211 \times 10^{-6}-7.4 \times 10^{-8}i & -6.25 \times 10^{-7}-9.6 \times 10^{-7}i \\
	0.00915+0.01638i & 0.0302+0.1268i & -0.0675-0.1885i
	\end{pmatrix}
	\end{equation*}
	
	\begin{equation*}\nonumber
	m_{S^{1/3}_1} = 6429.559~\mathrm{GeV},\,\quad m_{S^{1/3}_2} = 7268.734~\mathrm{GeV},\,\quad m_{S^{2/3}} = 7274.05~\mathrm{GeV}
	\end{equation*}
	
	\begin{equation*}\nonumber
	{\amber a_1} = 11.25~\mathrm{GeV},\,\quad \sin 2\theta = -1.384 \times 10^{-6}
	\end{equation*}
	\\
	\underline{Benchmark 7 ($\mathcal{B}7$):}
	
	\begin{equation*}\nonumber
	{\color{ForestGreen} \Upsilon} = 
	\begin{pmatrix}
	-7.144 \times 10^{-7}+2.881 \times 10^{-7}i & 0.000745-0.002317i & -2.867 \times 10^{-8}-3.669 \times 10^{-8}i \\
	1.177+0.526i & -3.373 \times 10^{-8}+3.89 \times 10^{-8}i & -0.0085-0.00562i \\
	-2.414 \times 10^{-7}-4.003 \times 10^{-7}i & -0.392+1.65i & 0.000523+0.0003847i
	\end{pmatrix}
	\end{equation*}
	
	\begin{equation*}\nonumber
	{\color{red} \Theta} = 
	\begin{pmatrix}
	0.001795-0.009334i & -6.771 \times 10^{-7}+1.165 \times 10^{-7}i & -3.044 \times 10^{-8}+3.031 \times 10^{-8}i \\
	-9 \times 10^{-5}+0.0001687i & -1.19 \times 10^{-8}-3.226 \times 10^{-7}i & 1.819+0.134i \\
	-0.01001-0.06185i & -0.00844-0.01543i & 0.2591+0.6186i
	\end{pmatrix}
	\end{equation*}
	
	\begin{equation*}\nonumber
	{\color{blue} \Omega} = 
	\begin{pmatrix}
	0.00449+0.02124i & 0.00143+0.04615i & -0.0178-0.1229i \\
	8.712 \times 10^{-7}+2.81 \times 10^{-8}i & 5.342 \times 10^{-6}-1.04 \times 10^{-7}i & -7.667 \times 10^{-7}-5.629 \times 10^{-7}i \\
	0.002432-0.008451i & -0.07037+0.05661i & 0.06006-0.02458i
	\end{pmatrix}
	\end{equation*}
	
	\begin{equation*}\nonumber
	m_{S^{1/3}_1} = 7586.233~\mathrm{GeV},\,\quad m_{S^{1/3}_2} = 8180.865~\mathrm{GeV},\,\quad m_{S^{2/3}} = 8186.574~\mathrm{GeV}
	\end{equation*}
	
	\begin{equation*}\nonumber
	{\amber a_1} = 16.36~\mathrm{GeV},\,\quad \sin 2\theta = -2.467 \times 10^{-6}
	\end{equation*}
	
	\cleardoublepage
	\bibliographystyle{JHEP}
	\bibliography{bib}
	
\end{document}